\documentclass[twoside]{article}

\usepackage{amssymb}
\usepackage{graphicx}
\usepackage{fullpage}
\usepackage{multirow}
\usepackage[round, authoryear]{natbib}
\usepackage{booktabs}
\usepackage{array}
\usepackage{url}
\usepackage{amsmath}
\usepackage{amstext}
\usepackage[lined,vlined,ruled,linesnumbered]{algorithm2e}
\usepackage{verbatim} 
\usepackage{etoolbox} 
\usepackage{lscape}
\usepackage{rotating}
\usepackage{hyperref}
\usepackage{subcaption}

\usepackage{comment}
\usepackage[x11names, rgb]{xcolor}
\usepackage[utf8]{inputenc}
\usepackage{tikz}
\usetikzlibrary{calc,arrows,shapes, fit, backgrounds, trees,hobby, positioning}

\pgfdeclarelayer{background}
\pgfdeclarelayer{foreground}
\pgfsetlayers{background,main,foreground}
\tikzstyle{every picture}+=[remember picture]

\tikzset{global scale/.style={
    scale=#1,
    every node/.style={scale=#1}
  }
}
\newcommand{\ra}[1]{\renewcommand{\arraystretch}{#1}}

\newtheorem{defi}{Definition}

\DeclareMathOperator*{\argmax}{arg\,max}

\DeclareMathAlphabet{\mathbbmsl}{U}{bbm}{m}{sl}

\usepackage{xcolor}

\usepackage{authblk}

\usepackage{pgfplots}
\usepgfplotslibrary{groupplots}
\pgfplotsset{compat=newest}

\usepackage{fancyhdr}
\newenvironment{customlegend}[1][]{%
    \begingroup
    \csname pgfplots@init@cleared@structures\endcsname
    \pgfplotsset{#1}%
}{%
    \csname pgfplots@createlegend\endcsname
    \endgroup
}%

\def\addlegendimage{\csname pgfplots@addlegendimage\endcsname}

\usepackage{wrapfig,epsf}

\title{\LARGE \bfseries Fast Detection of Community Structures using \\ Graph Traversal in Social Networks}
\author[1]{{Partha Basuchowdhuri}\thanks{Corresponding Author: \texttt{parthabasu.chowdhuri@heritageit.edu}}}
\author[2]{{Satyaki Sikdar\thanks{The work was done when the author was at Heritage Institute of Technology}}}
\author[1]{{Varsha Nagarajan}}
\author[1]{{Khusbu Mishra}}
\author[1]{\\Surabhi Gupta}
\author[1]{{Subhashis Majumder}}
\affil[1]{Department of Computer Science and Engineering,\protect\\Heritage Institute of Technology, Kolkata, WB, India}
\affil[2]{Department of Computer Science and Engineering,\protect\\ University of Notre Dame, Notre Dame, IN, USA}

\title{\LARGE \bfseries Fast Detection of Community Structures using \\ Graph Traversal in Social Networks}

\begin{document}
\label{firstpage}

\date{}

\maketitle

\begin{abstract}
Finding community structures in social networks is considered to be a challenging task as many of the proposed algorithms are computationally expensive and does not scale well for large graphs. Most of the community detection algorithms proposed till date are unsuitable for applications that would require detection of communities in real-time, especially for massive networks. The Louvain method, which uses \emph{modularity maximization} to detect clusters, is usually considered to be one of the fastest community detection algorithms even without any provable bound on its running time. We propose a novel graph traversal-based community detection framework, which not only runs faster than the Louvain method but also generates clusters of better quality for most of the benchmark datasets. We show that our algorithms run in $O(|V| + |E|)$ time to create an initial cover before using modularity maximization to get the final cover.
 
\end{abstract}

\section{Introduction}
Networks can be realized by a graph data structure defined by $G=(V, E)$, where $V$ denotes the vertex set, $E$ denotes the edge set and $n = |V|$, $m = |E|$. These networks exhibit certain implicit characteristics like community structures. \textit{Communities} in a network represent groups of vertices having dense intra-connections but sparse inter-connections. In this paper, we use the terms \emph{cluster} and \emph{community} interchangeably.
In a social network, a basic assumption is that the shortest paths are used to propagate information. Communities are connected via nodes, termed as \emph{brokers}, that are present on a large number of shortest paths among pairs of nodes in the network and hence, control the spread of information between communities. These nodes mark the boundary of communities. In our algorithm, we find such broker nodes that help to reach a new community and then spread the information among its members. In the process, we identify the nodes that are influenced by the brokers and also exhibit a high probability of belonging to the same community.

Most of the existing community detection algorithms involve lots of computations and hence are time-consuming. In 2002, in one of the early attempts to find communities from social networks, Girvan and Newman~\citep{GirvanandNewman2002} proposed an algorithm to detect hierarchical communities using repeated occurrence of an edge in all pair shortest paths as a metric. It was termed as the \emph{brokerage} value of an edge. This is one of the first papers that points out the significance of finding broker nodes or edges for detecting communities in social networks. This algorithm has a very high computational demand even for the sparse networks. Likewise, most of the other community detection algorithms are also computationally expensive and some of them suffer from the major disadvantage of detecting only disjoint communities. However, real world networks are generally found to have overlapping communities. Some algorithms take the number of communities required as an input, but, it may not always be possible to estimate the number of clusters without prior analysis of the network, if someone wants to find the best set of clusters.

In our paper, we try to overcome most of these existing drawbacks found in the known algorithms~\citep{fortunato2016}. Our algorithm uses popular traversal techniques like depth first traversal and breadth first traversal
and proves to perform considerably better and faster than other existing algorithms.


\section{Prior Works}

In this section, we mainly focus on the community detection algorithms known for having lesser running time. OPTICS~\citep{OPTICS}, a density based clustering works well with the benchmark datasets. It chooses an outlier point to start the algorithm and then traverses through the points to draw a plot to mark the denser and sparser regions and thereby detecting the clusters. We have guided our algorithm in a similar manner using a graph structure but have achieved a linear running time in the process. An overlapping community detection algorithm, ONDOCS~\citep{Chenetal2009}, takes help of visualization like OPTICS. It orders the nodes on the basis of their \textit{reachability scores}, which helps the user to understand the emerging network structure. After initial visualization, selected parameters are used for extracting communities, hubs, outliers from the network. It finds overlapping communities in a network with a worst case running time of $O(n \log n)$.

A modified version of overlapping Girvan-Newman (GN) algorithm~\citep{Gregory2008} was proposed to detect overlapping communities on the basis of a local form of betweenness. It discovers small-diameter communities in large networks and has a time complexity of $O(n \log n)$ for sparse networks. In one of his seminal papers, Mark Newman presented the notion of modularity $(Q)$~\citep{Newman06} of a clustering or a cover in a network, using a concept of minimization of inter-cluster edges and maximization of intra-cluster edges. Modularity, equipped with mathematical versatility, was very popularly used by the researchers in measuring goodness of covers and related topics. It opened up a new problem, popularly known as the modularity maximization problem. The decision problem corresponding to this optimization problem was later proved to be NP-complete~\citep{Brandes2006} and it is well accepted that heuristics can provide reasonably good solutions to the problem~\citep{Good2010}. Newman himself presented a solution~\citep{Newman04} to the problem but it was computationally expensive and practically infeasible for massive graphs. Later, as an extension of that work, a disjoint community detection technique was proposed by Clauset, Newman and Moore, now popularly known as CNM~\citep{CNM04}. It is essentially a greedy algorithm that uses efficient data structures to store and find the maximum gain in modularity incrementally and eventually finds communities in sub-quadratic running time. Another greedy modularity maximization algorithm by Blondel et. al.~\citep{BGLL08}, used a simplistic approach of looking into the neighbors of a node to look for increase in modularity. After deciding on a tentative split, it shrinks the network, thereby drastically reducing future computation. This method is popularly known as the Louvain method. Local notion of modularity has also been used for detection of communities by modifying the equation of modularity and including a parameter to address the resolution limit~\citep{xiang2016}.

Raghavan et. al. proposed a fast community detection algorithm~\citep{LPA} popularly known as the label propagation algorithm (LPA). It runs multiple breadth-first searches in successive iterations in a random manner such that labels propagate locally and after a few iterations converge to provide a stable final cover. This algorithm has a running time $O(m+n)$. The algorithm has several disadvantages - for example, it searches for the similar nodes locally by spreading labels to adjacent nodes and it needs multiple iterations of breadth-first traversals. The convergence of node labels can be mathematically guaranteed but it is not known if the number of iterations needed for the convergence of the node labels is dependent on $n$, $m$ or some other network parameter. Another community detection algorithm that claims to work fast in practice is Infomap~\citep{Infomap}. In this algorithm, the problem of community detection has been transformed into compression of information during its flow in the network. The algorithm uses random walks to move within the network and uses entropy-based information compression policies to find out the final cover. More recently, another linear time community detection technique (CGA) was presented by Yu Wang et. al.~\citep{Wangetal.2010} that detects communities in social networks taking into account information diffusion. It detects community structure in social networks using an approach of label propagation with a worst case running time of $O(m)$. Although it has a provable bound, there was no empirical proof to justify that it runs faster than the Louvain method. Also, their results could not be reproduced due to unavailability of any public release from the authors of the papers. The theoretical background of our method is different from these algorithms or methods but it has a similar bound for its running time. Another community detection technique that uses depth first traversal is LexDFS~\citep{creusefond2017}. The worst case time complexity of LexDFS algorithm has been reported to be $O(n \log n)$. We pit the performances of our method against the popular fast community detection techniques with publicly available releases. The Louvain method is widely accepted as the present state-of-the art in terms of finding disjoint communities from a network. Therefore, any improvement on the results obtained from the Louvain method can be considered as an improvement of the state-of-the-art. If Louvain method is started from a bias (i.e., a cover is fed as an input) instead of starting from the original network, the method can be faster and the structure of the final cover will largely depend on the initial bias. Therefore, a cover generated from a given bias might be quite different from the final cover generated by the Louvain method when it is run on the original network. This motivates us to generate a bias such that we can generate final covers that are better in quality.

Recently, local searches have been used to look into the neighborhood of a vertex to find the best possible community for that vertex~\citep{cui2014}. A new area of classifying nodes and thereby predicting communities is node and community embedding~\citep{zheng2016, wang2017, 7952933}. Usually node embedding outputs a vector representation for each node in the graph, such that two nodes being ``close'' on the graph have similar vector representations. Another recent method has used Jaccard co-efficient to find communities~\citep{meghanathan2016}. In this paper, Jaccard co-efficient has been used as a measure to detect locally dense group of nodes as the metric finds similarity between two adjacent nodes by detecting common neighbors between them. Similar to GN method, it detects communities by successive removal of edges ordered on the basis of non-decreasing values of Jaccard co-efficient.

\vspace*{-0.25cm}
\section{Problem Formulation}
\label{sec:prob}
A vertex $v$ is said to be influenced if a piece of information spreads to it from its neighbors (referred to as $\Gamma(v)$). Evidently, this is a temporal feature of the nodes and we assume that if a node $u$ gets influenced at time $t = i$, it remains influenced thereon, and all its uninfluenced neighbors get influenced at time $t = i + 1$. This is similar to applying the breadth-first traversal algorithm in a graph. 

\begin{defi}
$\bf{Influence}.$ A node $v \in V$ is said to be \emph{influenced}, once it has been discovered by using a graph traversal method.
\end{defi}

\begin{defi}
$\bf{Influenced}$ $\bf{Neighbors}$ $\bf{Score}.$ The \emph{influenced neighbors score} of a vertex $v$ at time $t=i$, $INS(v)_{t=i}$, is calculated as the fraction of the number of neighbors of vertex $v$ that have been influenced up to the previous time-stamp, i.e., $t=i-1$.
\end{defi}
\begin{equation}
INS(v) = \frac{\textit{Number of influenced neighbors of v}}{\textit{Degree(v)}}
\end{equation}

Clearly, $INS(v)$ lies between 0 and 1. For the starting node, it is zero. If all the neighbors of $v$ have been influenced before $v$ is processed, then $INS(v)$ will be 1. Initially INS value for all the nodes are unassigned. During the influence propagation, each node is accessed and its $INS$ value is calculated. Clearly, $INS$ value of a node is calculated only once, i.e., when it is discovered for the first time from the neighborhood of the node that is being processed. Therefore, calculation of $INS$ value of a node is dependent on the order of information spread (i.e., the traversal order), which, in turn, is dependent on the choice of the starting node.

\begin{figure}[ht]
	\centering
	\vspace{-1cm}
    \begin{tikzpicture}[every node/.style={draw,shape=circle, thick, node distance=1.5cm, fill=gray!25}, scale=0.75]

    \node (A)[ultra thick, fill=white] at (4, 4){A};
    \node (E) [ultra thick, below of=A, fill=white] {E};
    \node (B) [draw=gray, fill=black, text=white, right of=E] {B};
    \node (D) [left of=E] {D};
    \node (C) [ultra thick, below of=E] {C};
    \node (M) at (7, -1) {M};
    \node (L) [above right of=M] {L};
    \node (K) [below right of=M] {K};
    \node (N) [below right of=L] {N};
    \node (F) [ultra thick, fill=white] at (10, 2.5) {F};
    \node (G) [fill=white, above right of=F] {G};
    \node (I) [below right of=F] {I};
    \node (H) [fill=white, below right of=G] {H};
    
    \node[draw=none, fill=none] (GB) [right of=B, yshift=2cm] 
    {$\Gamma(B) = \{A, C, E, F\}$};

	\node[draw=none, fill=none] (INS)
	[below right of=GB, yshift=0.2cm, xshift=-0.8cm] {$INS(B) = \frac{1}{4} = 0.25$};

    \draw[ultra thick] (A) -- (B);
    \draw
    (A) -- (D)
    (A) -- (E);
    \draw[ultra thick] (B) -- (C)
    (B) -- (E)
    (B) -- (F);
    \draw
    (C) -- (D)
    (C) -- (M)
    (D) -- (E)
    (M) -- (L)
    (M) -- (K)
    (L) -- (I)
    (L) -- (K)
    (L) -- (N)
    (F) -- (G)
    (F) -- (I)
    (F) -- (H)
    (G) -- (I)
    (G) -- (H)
    (I) -- (H);
\end{tikzpicture}
    \caption{Calculating INS value of a node. The figure shows how $INS(B)$ is calculated.}
    \label{fig:ins}
\end{figure}
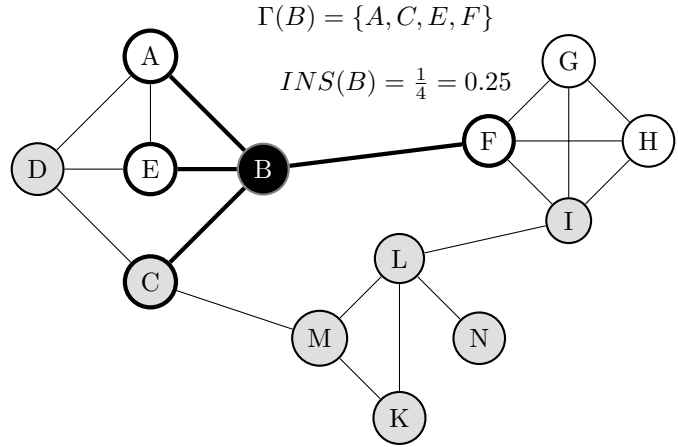

If $t$ is not explicitly mentioned for $INS(v)$, then it means $INS$ value has to be calculated for the current time-stamp. We explain how to calculate INS value of a node with an example shown in Figure ~\ref{fig:ins}. In this figure, we see that the $INS$ value for node B is being calculated. At the time of calculating $INS(B)$, a few nodes have already been influenced. In Figure ~\ref{fig:ins}, we can see that C, D, I, K L, M and N have been influenced. They have been shown in grey color in order to categorize them differently from the other nodes. If $INS$ value of B is being calculated at $t$=$i$, then the nodes in grey have been influenced at any time-stamp between time-stamps $t=1$ to $i-1$. At the time of calculating INS(B), we see how many nodes are in the neighborhood of B and how many of them are already influenced. C is the only influenced neighbor of B out of its four neighbors \{A, C, E, F\}, therefore $INS(B)$ is calculated to be $\frac{1}{4}$.

The intuition behind the definition of $INS$ comes from the fact that if two neighboring nodes are in same community and they are highly likely to have lots of common neighbors. During the influence propagation, if one of them is reached first, the influence will reach the other node and the common neighbors in the next time-stamp. If the other node is processed in the next time-stamp then it will find that many of its neighbors have already been influenced. It gives us an idea that the node is in a closely knit community and the information it wanted to propagate is already with many of its neighbors. The idea is intuitive and was first mentioned by Granovetter~\citep{Gran73}. Here, we present a mathematical form to use it in our proposed method.

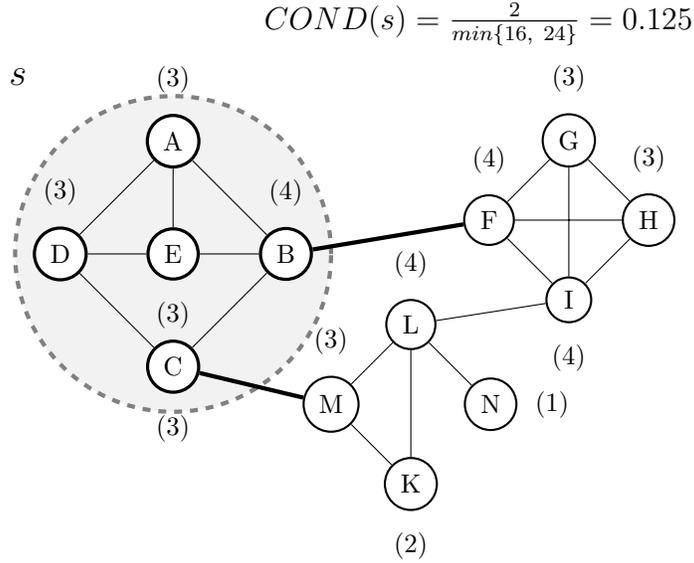
\begin{figure}[ht]
	\centering
	\vspace{-2cm}
	\begin{tikzpicture}[scale=0.7] 
\tikzstyle{every node}=[draw, fill=white, shape=circle, thick, node distance=1.5cm];
\node (A)[very thick, label={above: (3)}] at (4, 4){A};
\node (E) [very thick, below of=A, label={below: (3)}] {E};
\node (B) [very thick, right of=E, label={above: (4)}] {B};
\node (D) [very thick, left of=E, label={above: (3)}] {D};
\node (C) [very thick, below of=E, label={below: (3)}] {C};

\node (M) [label={above: (3)}] at (7, -1) {M};
\node (L) [above right of=M, label={above: (4)}] {L};
\node (K) [below right of=M, label={below: (2)}] {K};
\node (N) [below right of=L, label={east: (1)}] {N};

\node (F) [label={above: (4)}] at (10, 2.5) {F};
\node (G) [above right of=F, label={above: (3)}] {G};
\node (I) [below right of=F, label={below: (4)}] {I};
\node (H) [below right of=G, label={above: (3)}] {H};

%
\node[draw=none, fill=none] (INS)
[above right of=B, yshift=2cm, xshift=1.5cm] {\large $COND(s) = \frac{2}{min\{16,\ 24\}} = 0.125$};

\node (S) [draw=none, fill=none, above left of=A, yshift=-0.2cm, xshift=-1cm] {{\Large $s$}};  

\draw[ultra thick] (B) -- (F) (C) -- (M);

\draw
(A) -- (B)
(A) -- (D)
(A) -- (E)
(B) -- (C)
(B) -- (E)
(C) -- (D)
(D) -- (E)
(M) -- (L)
(M) -- (K)
(L) -- (I)
(L) -- (K)
(L) -- (N)
(F) -- (G)
(F) -- (I)
(F) -- (H)
(G) -- (I)
(G) -- (H)
(I) -- (H);

\begin{pgfonlayer}{background}
	\draw[draw=gray, fill=gray!10, dashed, ultra thick] (E) circle(3);
\end{pgfonlayer}

\end{tikzpicture}

\caption{Calculating conductance of a cut set $s = \{A, B, C, D, E\}$. Degrees of the nodes are given in parentheses.}
\label{fig:cond}
\end{figure}

\begin{defi}
$\bf{Cut}.$ In a graph $G(V, E)$, \emph{cut} of a cluster $s$ is defined as the sum of the weight of edges from cluster $s$ to its complement $V \setminus V_s$~\citep{Whang2013}.
\end{defi}

\begin{defi}
$\bf{Conductance}.$  Let $V_s$ be the set of vertices in a cluster $s$. The \emph{conductance} of $s$ can be defined as the cut dividing the least number of edges incident on either the cluster or the rest of nodes in the network $(V \setminus V_{s})$. In other words, it is the probability of leaving the cluster by a one-hop walk starting from the smaller set between $V_{s}$ and $V \setminus V_{s}$~\citep{Whang2013}.
\end{defi}
The formulation of \emph{conductance} can be given by,
\begin{equation}
COND(s) = \frac{C(V_s, V \setminus V_s)}{min(C(V_s, V), C(V \setminus V_s, V))},
\label{eq:cond}
\end{equation}
where, $C(A, B)$ is defined as the sum of the weights of edges between subgraphs with node sets $A$ and $B$. Note that $A \cap B$ may not always be a null set.  Fig.~\ref{fig:cond} shows an example how conductance of a cut set $s = \{A, B, C, D, E\}$ is computed for the given graph.

\begin{defi}
$\bf{Broker}$ $\bf{Nodes}.$ A vertex $v$ is said to be a \emph{broker node}, if at present time-stamp, $INS(v)$ is less than a predefined threshold $r$, where $0 \leq r \leq 1$.
\end{defi}

The fraction $r$ is termed as \emph{community threshold} and may differ from one network to another for finding the best available community structure present in the network. Empirically, it was found (as shown in Table~\ref{tab:threshold}) that the variation of the quality of the communities obtained is marginal when $r$ lies in the range of $0.6$ to $0.8$. In a sparse graph, the community threshold for a network may have a smaller value than that of in a dense graph. The nodes that are not broker nodes are declared as \emph{community nodes}.


\begin{defi}
$\bf{Community}$ $\bf{Nodes}.$ A vertex $v$ is said to be a \emph{community node}, if at the present time-stamp or at $t=i$, $INS(v)$ is greater than or equal to a threshold $r$, where $0 \leq r \leq 1$.
\end{defi}

When using conductance as the objective function, broker nodes are defined as those nodes that when added to a cluster, cause an increase or no change in the conductance value. Nodes that are not brokers are declared as community nodes. These definitions have been repeatedly used in the methods we have proposed in this paper.

\subsection*{Problem Statement}
Given a graph $G=(V,E)$, where $n = |V|$ and $m = |E|$, find a cover (set) of $k$ partitions (communities) from the hypothesis space consisting of all possible covers, represented by $C = \{C_1, C_2, C_3,...\}$, where $C_i$ is the $i$-th cover from the hypothesis space, $V_i = \cup_{j=1}^k V_{ij}$ and $E_i = \cup_{j=1}^k E_{ij}$, such that the value of the clustering goodness measure is maximized over all such possible covers where $1 \le k \le n$.

Note that the goodness measure can be performed by any goodness function such as modularity~\citep{Newman06} or overlapping modularity~\citep{NPNB07,HWXJ09,Nicosia2009}.
In other words, for a graph $G$, if the value of the goodness function for cover $C_i$ is given by $Q(G, C_i)$, then our aim is to find
\begin{eqnarray*}
&\argmax_{c \in C} Q(G, c) = \{Q(G, C_i) &|\;\forall\, C_i, C_j \in C, Q(G, C_i) \geq Q(G, C_j)\}.
\end{eqnarray*}
The target is to find the $c$, for which $Q(G, c)$ is maximum out of all possible covers from the hypothesis space $C$. Cover $c$ is considered to be an overlapping partition if intersection of the vertex sets of any two clusters produces a set that is not null. Otherwise, the cover $c$ is considered to be a disjoint partition. The problem presented here is essentially a goodness maximization problem. The decision version of goodness maximization problem, with modularity being the function for measuring goodness, has been proved to be NP-complete~\citep{Brandes2006}.

For clarification, it should be noted that in our proposed method we try to achieve final clusters with good cluster quality. It is independent of any particular clustering goodness measure. We try to achieve high goodness measure without using any particular goodness measure as the objective function. Instead we use $INS(v)$ and $COND(s)$ as objective functions to maximize intra-cluster edges and minimize inter-cluster edges.

\section{Traversal-based Community Detection}

In this section, we propose a framework for two community detection methods which aim to find communities from a social network in linear running time (linear in terms of the size of the network) using traversal techniques. During the traversal, we first label nodes as brokers or community nodes based on an objective function. Next, we place the broker nodes in the community where most of its neighbors lie. We then reduce the graph on the basis of this initial split, where every cluster is merged into one super-vertex, with a weighted self-loop depicting the total number of intra-cluster edges. Finally, modularity maximization is run to generate the final cover. The method works as if a seed node is chosen to spread some information to the whole network. So the seed node spreads the information to all its neighbors and they in turn spread the information to their neighbors, who do not have the information yet. This process goes on iteratively and in the process we analyze the path of traversal to find out the communities. In this community detection algorithm, we use traversal techniques in sequential manner and thereby achieve a linear running time. We use the same framework to generate different clusters by devising two different methods on the basis of the objective functions mentioned in Section~\ref{sec:prob}.

\subsection{Part 1: Finding the Broker Nodes to Outline the Communities}
 The main idea of our method is to traverse the whole graph and partition it into a set of communities by observing how much of influence has reached a node's neighborhood or how connected its neighborhood is. As mentioned earlier in this paper, communities are subgraphs with dense intra-connections and sparse inter-connections. Broker nodes reside in the bordering areas of a community, i.e., the areas where communities overlap. As a result, a piece of information that is exclusive to one community can spread to an adjacent community only via the broker nodes. These broker nodes behave as transition points between two or more communities. When a piece of information reaches to a community, it first reaches the broker nodes and then spreads among its members. Then, through some other broker nodes the information is passed on to another adjacent community. In our method, we follow the pattern of information flow via the broker nodes to explore and thereby detect the community structure of a graph in the process.

\begin{flushleft}
\textbf{INS-based Algorithm:}
\end{flushleft}

 In this method, a node in a graph is identified as either a broker node or a community node on the basis of its $INS$ value. 
 When a node is encountered during the traversal, we find the number of its neighbors that are carrying the information during that time stamp. If the fraction of neighbors of a node $v$ carrying the information is less than $r$ (say 0.75) then we assume that the information has not yet reached the community $v$ belongs to, and $v$ is one of the first nodes from its community to receive the information. Therefore, $v$ is termed as a \textit{broker} node and it is marked with a community label, which is same as its node label. Otherwise, the node is categorized as a \textit{community} node.

\pagebreak
\begin{flushleft}
\textbf{Conductance-based Algorithm:}
\end{flushleft}

In this method, a node is identified as a broker when addition of the node in the presently growing cluster $s$ increases $COND(s)$. Initially, when the process starts from a single node, the cluster consists of a single node and $COND(s)$ is 1. As new nodes are picked up during the traversal, we try to include the newly reached node in the cluster and check how the conductance changes. If it increases then it is considered to be a broker node, else the process to grow the cluster continues. Intuitively, it should stop at the borders of the prospective clusters. Nodes, other than the broker nodes, are marked as community nodes.

The community nodes identified during the traversal from one broker node to the next identified broker node are said to belong to the same community as the brokers mark the border of two communities. The community label for a group of community nodes found subsequently during traversal is same as the label of the broker node that led to the traversal of those community nodes.

\subsection{Part 2: Allocating Broker Nodes to the Communities}
In this part, we put the broker nodes in the communities they are most likely to belong. This decision is made based on a metric called belonging probability that calculates the ratio between the number of edges that exists from a broker node to all the nodes in a community out of the maximum number of edges possible between the broker node and that community. The idea behind defining such a metric is to find out the community to which a broker node is connected to with most number of edges. The absolute value of the number of connections to a community may not correctly represent the association of a broker node to that community. Increase in size of communities leads to increase in the possible number of connections with the broker node. Hence, if the number of actual connections are normalized by the possible number of connections that can be made with a community, the probability of a broker node belonging to that community remains independent of the size of the community. Eventually, broker nodes are placed in the community that leads to the highest belonging probability.

Some of the broker nodes may still fall into communities they are not supposed to be a part of as identified by the ground truth. The sequential nature of the traversal-based detection of communities may lead to such a situation. We solve this problem by identifying the community in which most of the neighbors of a broker node lie. In case of a tie (i.e., if the cardinality of the set of neighbors in a particular community maximizes for more than one community), we do not assign a community label to the broker node but leave it to the modularity maximization step to merge it with one of the clusters on the basis of the topological structure. A similar policy is maintained for the broker nodes with no community node in its neighbor. In the first part of modularity maximization, we reduce the present clusters into super-vertices and thereby transform the network into an undirected network of super-vertices. Every super-vertex consists of a self-loop, which has a weight equivalent to twice the number of intra-cluster edges in that cluster. The edges between a pair of super-vertices have weights equivalent to the number of inter-cluster edges between those two clusters. This step is termed as the reduction of the network. After the reduction, modularity changes for potential merges are calculated and increase in modularity is greedily maximized. In this step, the super-vertices are merged iteratively to produce the final cover. Due to the nature of the algorithms, the initial clusters are expected to be fragmented, however, with high precision. Therefore, a modularity maximization process may merge those high precision community fragments to achieve communities with both high precision and high recall. Such communities are likely to be more similar to the ground truth than the initial fragments.

\subsection{Mathematical Definition of Community Nodes}

The following definition of conductance is equivalent to the one defined in equation (2). In this section, we also abbreviate $COND$ to $C$ for convenience.
\begin{displaymath}
C(S)=\frac{\sum_{i\in S,j\notin S}A_{ij}}{min\{k_{S},\ k_{\bar{{S}}}\}}
\end{displaymath}

where the numerator is the cut size, i.e., the number of edges from $S$ to $\bar{{S}}$, $k_{S}$ is the sum of degree of nodes in cluster $S$.

We are interested in the change in conductance of cluster $S$ triggered by the addition of a single neighboring node (called the target node) to the cluster. If the conductance decreases after the addition, the target node is classified as a \textbf{community} node and added to the community $S$, otherwise it is classified as a \textbf{broker} node.

Given a graph and a partial cluster $S$, we can classify the nodes in the graph into three types,
\begin{enumerate}
\item the nodes currently in the cluster $S$,
\item the target node $t$, which potentially could be added to $S$,
\item the other nodes $o$ which belongs to $O$ (i.e., $V-S-\{t\}$).
\end{enumerate}

Say, $k_{t}$ is the degree of the target node, $k_{S}$ and $k_{O}$ are the degrees of the clusters $S$ and $O$ respectively, $k_{t,S}$ is the number of the edges incident from the target $t$ to the cluster $S$, $\alpha$ is the number of edges incident from the cluster $S$ to the set $O$. In other words, $\alpha$ represents the number of edges in the cut set \textbf{not} incident on the target node $t$.

We investigate the numerator of the definition, i.e., the cut size. Notice that the cut size initially (before $t$ is added to $S$) is $\alpha+k_{t,S}$. After $t$ is added to $S$, the cut size becomes $\alpha+(k_{t}-k_{t,S})$. This is because when $t$ becomes
a part of $S$, it brings $k_{t}$ many edges with itself. Out of which only $k_{t}-k_{t,S}$ contribute to the cut size, since $k_{t,S}$ edges become part of the cluster $S$ after the merge.

For the denominator, initially it is $min\{k_{S},(k_{t}+k_{O})\}$. After $t$'s addition, it becomes $min\{(k_{S}+k_{t}),k_{O}\}$. Since
we pick the minimum of the two quantities, three cases arise:
\begin{enumerate}
\item $k_{S}<k_{t}+k_{O}$ and $k_{S}+k_{t}<k_{O}$,
\item $k_{S}<k_{t}+k_{O}$ and $k_{S}+k_{t}\ge k_{O}$,
\item $k_{S}\ge k_{t}+k_{O}$ (this guarantees that $k_{S}+k_{t}>k_{O}$, so there is no $4^{th}$ case).
\end{enumerate}

The initial conductance before $t$ is added to $S$ is represented as $C_{old}$ and after $t$'s addition it is $C_{new}$. For each
case, we find the condition on $k_{t,S}.$

\textbf{Case I: $k_{S}<k_{t}+k_{O}$ and $k_{S}+k_{t}<k_{O}$}

\begin{align*}
C_{old} & = \frac{\alpha + k_{t,S}}{k_{S}}, C_{new} = \frac{\alpha + k_{t} - k_{t,S}}{k_{S} + k_{t}}\\
C_{old} - C_{new} & = \frac{\alpha+k_{t,S}}{k_{S}} - \frac{\alpha + k_{t}-k_{t,S}}{k_{S} + k_{t}}\\
k_{S} \cdot (k_{S} + k_{t}) \cdot (C_{old} - C_{new}) & =(\alpha + k_{t,S}) \cdot (k_{S} + k_{t}) - k_{S} \cdot (\alpha + k_{t} - k_{t,S})\\
 & =\alpha \cdot k_{S} + \alpha \cdot  k_{t} + k_{S} \cdot k_{t,S} + k_{t} \cdot k_{t,S} - \alpha \cdot k_{S} - k_{S} \cdot k_{t} + k_{S} \cdot k_{t,S}\\
 & = k_{t,S} \cdot (2k_{S} + k_{t}) + k_{t} \cdot (\alpha - k_{S})
\end{align*}

For $t$ to be a community node, $C_{old} - C_{new} > 0$. Note that $k_{S}(k_{S}+k_{t})$ is always positive. Therefore,

\begin{align}
k_{t,S}\cdot(2k_{S} + k_{t}) & + k_{t}\cdot(\alpha - k_{S}) > 0\nonumber \\
k_{t,S}\cdot(2k_{S} + k_{t}) & > k_{t}\cdot(k_{S} - \alpha)\nonumber \\
k_{t,S} & > \frac{k_{t}\cdot(k_{S} - \alpha)}{2k_{S} + k_{t}}
\label{eqn:cond1}
\end{align}

\pagebreak
\textbf{Case II: $k_{S}<k_{t}+k_{O}$ and $k_{S}+k_{t}\ge k_{O}$}

\begin{align*}
C_{old} &  = \frac{\alpha + k_{t,S}} \cdot {k_{S}}, C_{new} = \frac{\alpha + k_{t} - k_{t,S}}{k_{O}}\\
C_{old} - C_{new} & = \frac{\alpha + k_{t,S}}{k_{S}} - \frac{\alpha + k_{t} - k_{t,S}}{k_{O}}\\
k_{S} \cdot k_{O}(C_{old} - C_{new}) & = (\alpha + k_{t,S}) \cdot k_{O} - k_{S} \cdot (\alpha + k_{t} - k_{t,S})\\
 & = \alpha \cdot k_{O} + k_{O} \cdot k_{t,S} - \alpha \cdot k_{S} - k_{S} \cdot k_{t} + k_{S} \cdot k_{t,S}\\
 & = k_{t,S} \cdot (k_{O} + k_{S}) + \alpha \cdot (k_{O} - k_{S}) - k_{S} \cdot k_{t}
\end{align*}

For $t$ to be a community node, $C_{old}-C_{new}>0$. Note that $k_{S}k_{O}$ is always positive. Therefore,

\begin{align}
k_{t,S} \cdot (k_{O}+k_{S}) & +\alpha \cdot (k_{O}-k_{S})-k_{S} \cdot k_{t} > 0\nonumber \\
k_{t,S} \cdot (k_{O}+k_{S}) & > k_{S} \cdot k_{t}+\alpha \cdot (k_{S}-k_{O})\nonumber \\
k_{t,S} & > \frac{k_{S} \cdot k_{t}+\alpha \cdot (k_{S}-k_{O})}{(k_{S}+k_{O})}
\label{eqn:cond2}
\end{align}

\textbf{Case III: \textmd{\normalsize{}$k_{S}\ge k_{t}+k_{O}$}}

\begin{align*}
C_{old} & =\frac{\alpha+k_{t,S}}{k_{t}+k_{O}}, C_{new} =\frac{\alpha+k_{t}-k_{t,S}}{k_{O}}\\
C_{old} - C_{new} & =\frac{\alpha + k_{t,S}}{k_{t} + k_{O}} - \frac{\alpha + k_{t} - k_{t,S}}{k_{O}}\\
k_{O} \cdot (k_{O} + k_{t}) \cdot (C_{old} - C_{new}) & =(\alpha + k_{t,S}) \cdot k_{O} - (k_{t} + k_{O}) \cdot (\alpha + k_{t} - k_{t,S})\\
 & = \alpha \cdot k_{O} + k_{O} \cdot k_{t,S} - \alpha \cdot k_{t} - k_{t}^{2} + k_{t} \cdot k_{t,S} - \alpha \cdot k_{O} - k_{O} \cdot k_{t} + k_{O} \cdot k_{t,S}\\
 & = k_{t,S} \cdot (2k_{O} + k_{t}) - k_{t} \cdot (\alpha+k_{t}+k_{O})
\end{align*}

For $t$ to be a community node, $C_{old}-C_{new}>0$. Note that $k_{O}(k_{O}+k_{t})$ is always positive. Therefore,

\begin{align}
k_{t,S} \cdot (2k_{O} + k_{t}) & - k_{t} \cdot (\alpha + k_{t} + k_{O}) > 0\nonumber \\
k_{t,S} \cdot (2k_{O} + k_{t}) & > k_{t} \cdot (\alpha + k_{t} + k_{O})\nonumber \\
k_{t,S} & > \frac{k_{t} \cdot (\alpha + k_{t} + k_{O})}{2k_{O} + k_{t}}
\label{eqn:cond3}
\end{align}

Equations~\ref{eqn:cond1}, \ref{eqn:cond2} and \ref{eqn:cond3} give the conditions for classifying the target node to be a community node. Combining them together we can say,

\begin{equation}
k_{t,S} >
\begin{cases}
\frac{k_{t} \cdot (k_{S}-\alpha)}{2k_{S} + k_{t}} & \text{if $k_{S}<k_{t}+k_{O}$ and $k_{S}+k_{t}<k_{O}$}\\
\frac{k_{S} \cdot k_{t}+\alpha \cdot (k_{S}-k_{O})}{(k_{S}+k_{O})} & \text{if $k_{S}<k_{t}+k_{O}$ and $k_{S}+k_{t}\ge k_{O}$} \\
\frac{k_{t} \cdot (\alpha+k_{t}+k_{O})}{2k_{O} + k_{t}} &\text{if $k_{S} > k_{t}+k_{O}$}
\end{cases}
\label{eqn:cond}
\end{equation}

\subsection{Illustrating Examples - Classifying Community and Broker Nodes}

\subsubsection{Example 1}

\begin{figure}[ht]
{
    \vspace{0.5cm}
    \centering
    \begin{tikzpicture}[scale=0.75] 
\tikzstyle{every node}=[draw, fill=white, shape=circle, thick, node distance=1.5cm, opacity=0.35];
\node (A)[
] at (4, 4){A};

\node (E) [
below of=A, 
] {E};

\node (B) [
right of=E, 
] {B};

\node (D) [
left of=E, 
] {D};
\node (C) [
below of=E,
] {C};

\node (M) [
opacity=0.7,
] at (7, -1) {M};

\node (L) [
opacity=1, 
above right of=M, 
label={above: (4)}
] {L};

\node (K) [
opacity=0.7,
below right of=M, 
] {K};

\node (N) [
opacity=1, 
below right of=L, 
label={east: (1)}
] {N};

\node (F) [
] at (10, 2.5) {F};

\node (G) [
above right of=F, 
] {G};

\node (I) [
opacity=0.7,
below right of=F, 
] {I};

\node (H) [
below right of=G, 
] {H};

\node (S) [draw=none, fill=none, opacity=1, above right of=N, yshift=-0.2cm, xshift=0.7cm] {{\Large $S = \{L, N\}$}};  

\draw[ultra thick] (L) -- (I)
(L) -- (K)
(L) -- (M);

\draw[thick] (L) -- (N);
 
\draw[opacity=0.35]
(A) -- (B)
(A) -- (D)
(A) -- (E)
(B) -- (C)
(B) -- (E)
(B) -- (F)
(C) -- (M)
(C) -- (D)
(D) -- (E)
(M) -- (K)
(F) -- (G)
(F) -- (I)
(F) -- (H)
(G) -- (I)
(G) -- (H)
(I) -- (H);

\begin{pgfonlayer}{background}
    \draw[black, fill=gray!15, dashed](L.north west) to[closed,curve through={(L.north) .. (L.north east) ..  (N.east) .. (N.south)}] (L.west);
\end{pgfonlayer}

\end{tikzpicture}
    \caption{The graph with the cluster $S = \{L, N\}$. The cut edges are drawn with thick lines. The number in parentheses near the nodes represent the degree.}
    \label{fig:cond1}
}
\end{figure}
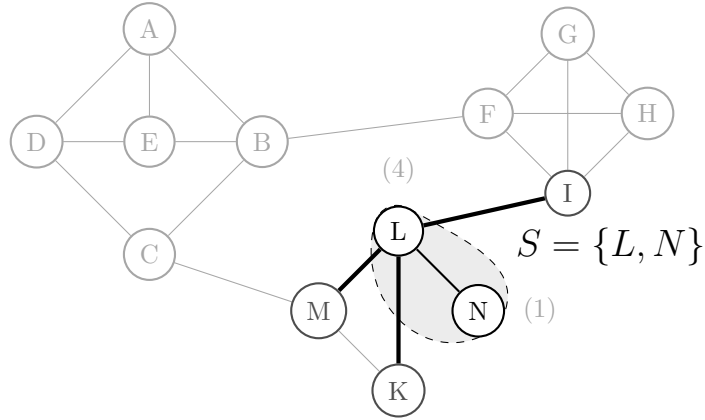

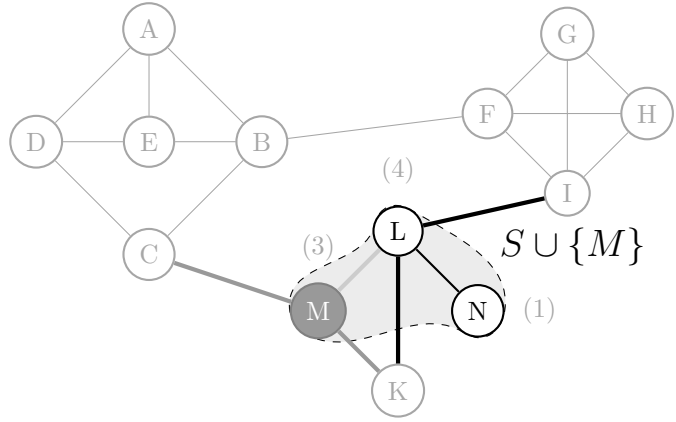
\begin{figure}[ht]
{
    \centering
    \begin{tikzpicture}[scale=0.75] 
\tikzstyle{every node}=[draw, fill=white, shape=circle, thick, node distance=1.5cm, opacity=0.35];
\node (A)[
] at (4, 4){A};

\node (E) [
below of=A, 
] {E};

\node (B) [
right of=E, 
] {B};

\node (D) [
left of=E, 
] {D};
\node (C) [
below of=E,
] {C};

\node (M) [
opacity=1, 
fill=black!40,
draw=gray,
text=white,
label={above: (3)}
] at (7, -1) {M};

\node (L) [
opacity=1, 
above right of=M, 
label={above: (4)}
] {L};

\node (K) [
below right of=M, 
] {K};

\node (N) [
opacity=1, 
below right of=L, 
label={east: (1)}
] {N};

\node (F) [
] at (10, 2.5) {F};

\node (G) [
above right of=F, 
] {G};

\node (I) [
below right of=F, 
] {I};

\node (H) [
below right of=G, 
] {H};

\node (S) [draw=none, fill=none, opacity=1, above right of=N, yshift=-0.2cm, xshift=0.2cm] {{\Large $S \cup \{M\}$}};  

\draw[draw=black, ultra thick] 
(L) -- (I)
(L) -- (K);
\draw[draw=black!20, 
ultra thick] 
(L) -- (M);

\draw[thick] (L) -- (N);
 
\draw[draw=black!40, 
ultra thick] 
(M) -- (C)
(M) -- (K);
 
\draw[opacity=0.35]
(A) -- (B)
(A) -- (D)
(A) -- (E)
(B) -- (C)
(B) -- (E)
(B) -- (F)
(C) -- (D)
(D) -- (E)
(F) -- (G)
(F) -- (I)
(F) -- (H)
(G) -- (I)
(G) -- (H)
(I) -- (H);

\begin{pgfonlayer}{background}
    \draw[black, fill=gray!15, dashed](L.north west) to[closed,curve through={(L.north) .. (L.north east) ..  (N.east) .. (N.south) .. (N.south west) .. (M.south) .. (M.west) }] (L.west);
\end{pgfonlayer}
\end{tikzpicture}
    \caption{$M$ is the target node. The number in parentheses above the nodes represent the degree of the nodes. The edges contributing to $k_{t,S},\ (k_t - k_{t,S})$ and $\alpha$  are shown in light gray, gray and black respectively.}
    \label{fig:cond2}
}
\end{figure}

Figure~\ref{fig:cond1} and \ref{fig:cond2} have been used to illustrate the first example. From Figure~\ref{fig:cond1}, we can calculate the parameters needed to classify the target node $M$ as a community node or a broker node. The parameters, in this example, are as follows
$$
S = \{L, N\},\ k_S = 5,\ t = M, \ k_t = 3,\ k_{t, S} = 1,\ \alpha = 2,\ k_O = 32.
$$
We have $k_S < k_t + k_O$ and $k_S + k_t < k_O$. Thus, we check the condition for case I to check if $M$ is a community node.

\begin{equation*}
    k_{t,S} > \frac{k_t \cdot (k_S - \alpha)}{2k_S + k_t}
\end{equation*}

Here, $LHS = k_{t,S} = 1$. Plugging in the values in the RHS, we obtain,
\begin{align*}
    RHS &= \frac{k_t \cdot (k_S - \alpha)}{2k_S + k_t} \\
    &= \frac{3 \cdot (5 - 2)}{2 \cdot 5 + 3} \\
    RHS &= \frac{9}{13}
\end{align*}

Thus, we have LHS equal to RHS. From Eq. 6, we can say, $I$ is \textbf{not} a community node,and is therefore added to $S$ (as shown in Figure~\ref{fig:cond2}).

\subsubsection{Example 2}

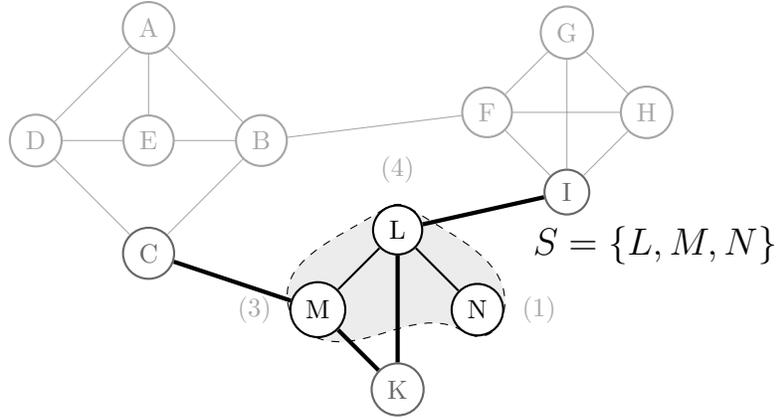
\begin{figure}[ht]
{
    \centering
    \begin{tikzpicture}[scale=0.75] 
\tikzstyle{every node}=[draw, fill=white, shape=circle, thick, node distance=1.5cm, opacity=0.35];
\node (A)[
] at (4, 4){A};

\node (E) [
below of=A, 
] {E};

\node (B) [
right of=E, 
] {B};

\node (D) [
left of=E, 
] {D};
\node (C) [
below of=E,
opacity=0.6,
] {C};

\node (M) [
opacity=1,
fill=white,
label={left: (3)}
] at (7, -1) {M};

\node (L) [
opacity=1, 
above right of=M, 
label={above: (4)}
] {L};

\node (K) [
opacity=0.6,
below right of=M, 
] {K};

\node (N) [
opacity=1, 
below right of=L, 
label={east: (1)}
] {N};

\node (F) [
] at (10, 2.5) {F};

\node (G) [
above right of=F, 
] {G};

\node (I) [
opacity=0.6,
below right of=F, 
] {I};

\node (H) [
below right of=G, 
] {H};

\node (S) [draw=none, fill=none, opacity=1, above right of=N, yshift=-0.2cm, xshift=1.3cm] {{\Large $S = \{L, M, N\}$}};  

\draw[ultra thick] 
(L) -- (I)
(L) -- (K)
(M) -- (K)
(C) -- (M);

\draw[thick] 
(L) -- (N)
(L) -- (M);
 
\draw[opacity=0.35]
(A) -- (B)
(A) -- (D)
(A) -- (E)
(B) -- (C)
(B) -- (E)
(B) -- (F)
(C) -- (D)
(D) -- (E)
(F) -- (G)
(F) -- (I)
(F) -- (H)
(G) -- (I)
(G) -- (H)
(I) -- (H);

\begin{pgfonlayer}{background}
    \draw[black, fill=gray!15, dashed](L.north west) to[closed,curve through={(L.north)  .. (L.north east) .. (N.east) .. (N.south) .. (N.south west) ..  (M.south) .. (M.west)}] (L.north west);
\end{pgfonlayer}

\end{tikzpicture}
    \caption{The graph with the cluster $S = \{L, M, N\}$. The cut edges are drawn with thick lines. The number in parentheses near the nodes represent the degree.}
    \label{fig:cond3}
}
\end{figure}

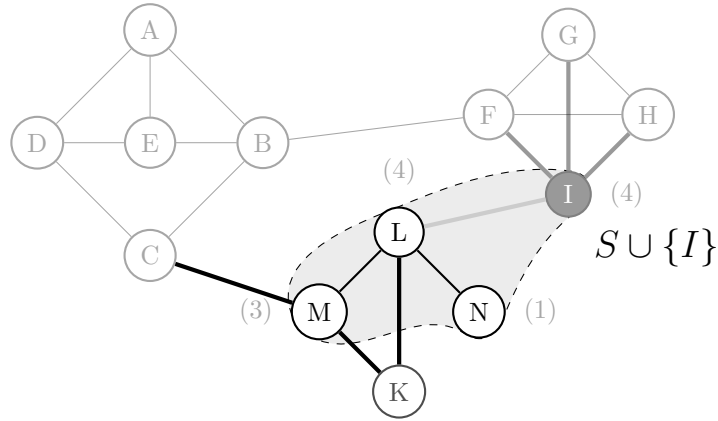
\begin{figure}[ht]
{
    \centering
    \begin{tikzpicture}[scale=0.75] 
\tikzstyle{every node}=[draw, fill=white, shape=circle, thick, node distance=1.5cm, opacity=0.35];
\node (A)[
] at (4, 4){A};

\node (E) [
below of=A, 
] {E};

\node (B) [
right of=E, 
] {B};

\node (D) [
left of=E, 
] {D};
\node (C) [
below of=E,
] {C};

\node (M) [
opacity=1,
fill=white,
label={left: (3)}
] at (7, -1) {M};

\node (L) [
opacity=1, 
above right of=M, 
label={above: (4)}
] {L};

\node (K) [
opacity=0.7,
below right of=M, 
] {K};

\node (N) [
opacity=1, 
below right of=L, 
label={east: (1)}
] {N};

\node (F) [
] at (10, 2.5) {F};

\node (G) [
above right of=F, 
] {G};

\node (I) [
fill=black!40,
draw=gray,
text=white,
opacity=1,
below right of=F, 
label={right: (4)}
] {I};

\node (H) [
below right of=G, 
] {H};

\node (S) [draw=none, fill=none, opacity=1, above right of=N, yshift=-0.2cm, xshift=1.3cm] {{\Large $S \cup \{I\}$}};  

\draw[ultra thick, draw=black!20]
(L) -- (I);

\draw[draw=black, ultra thick] 
(M) -- (K)
(M) -- (C)
(L) -- (K);

\draw[thick] 
(L) -- (N)
(L) -- (M);

\draw[ultra thick, draw=black!40]
(I) -- (F)
(I) -- (G)
(I) -- (H);
 
\draw[opacity=0.35]
(A) -- (B)
(A) -- (D)
(A) -- (E)
(B) -- (C)
(B) -- (E)
(B) -- (F)
(C) -- (M)
(C) -- (D)
(D) -- (E)
(M) -- (K)
(F) -- (G)
(F) -- (H)
(G) -- (H);

\begin{pgfonlayer}{background}
    \draw[black, fill=gray!15, dashed](L.north west) to[closed,curve through={(L.north)  .. (I.north) .. (I.north east) .. (I.south) ..  (N.east) .. (N.south) .. (N.south west) ..  (M.south) .. (M.west)}] (L.north west);
\end{pgfonlayer}

\end{tikzpicture}
    \caption{$I$ is the target node. The number in parentheses above the nodes represent the degree of the nodes. The edges contributing to $k_{t,S},\ (k_t - k_{t,S})$ and $\alpha$ are shown in light gray, gray and black respectively.}
    \label{fig:cond4}
}
\end{figure}

Figures 5 and 6 illustrate the second example. From Fig. 3, we can calculate the parameters needed to classify the target node $I$. The parameters, in this example, are as follows: 
$$S = \{L, M, N\}, k_S = 8, t = I, k_t=4, k_{t, S} = 1, \alpha = 3, k_O = 28. $$

We have $k_S < k_t + k_O$ and $k_S + k_t < k_O$. Thus, we check the condition for case I to check if $I$ is a community node.

\begin{equation*}
    k_{t,S} > \frac{k_t \cdot (k_S - \alpha)}{2k_S + k_t}
\end{equation*}
Here, $LHS = k_{t,S} = 1$. Plugging in the values in the RHS, we obtain,
\begin{align*}
    RHS &= \frac{k_t \cdot (k_S - \alpha)}{2k_S + k_t} \\
    &= \frac{4 \cdot (8 - 3)}{2 \cdot 8 + 4} \\
    RHS &= \frac{4 \cdot 5}{16 + 4} = 1
\end{align*}

Thus, we have $LHS$ equal to $RHS$. From Equation~\ref{eqn:cond} we can say, $M$ is \textbf{not} a community node, but a \textbf{broker} and is therefore not  added to $S$.

\subsection{Proposed Algorithms}

We have described our proposed method in algorithms~\ref{algo:lcd},~\ref{algo:nodecatins},~\ref{algo:bfs},~\ref{algo:ins},~\ref{algo:cond},~\ref{algo:postprocess},~\ref{algo:reduce},~\ref{algo:modmax} and ~\ref{algo:nodecatcond}, incrementally. Algorithm~\ref{algo:lcd} describes the central LINCOM method algorithm replaced by method, where a broker stack $S$ and a community queue $Q$ have been used as primary data structures to facilitate the traversal. We call other procedures from Algorithm~\ref{algo:lcd} to update $S$ and $Q$ and detect communities in the process. The phenomenon termed as ``spread of information'' has been described by Algorithms~\ref{algo:bfs} and ~\ref{algo:nodecatins}. The process of spread imitates breadth-first traversal. As a part of the data structures used, we have also used three node attributes to store flags for all the nodes, regarding whether it is a broker node or a community node (using nodeType), whether it has been covered during the traversal or not (using covered) and which community it presently belongs to (using community).

Algorithm~\ref{algo:nodecatins} describes a procedure NODE-CAT-INS that is called by LINCOM (when using INS), until all the nodes in the graph have been categorized. In Algorithm~\ref{algo:lcd}, we call a generic version of the procedure referred to as NODE-CAT. But while presenting the pseudocodes, we have named the INS-based and COND-based versions of NODE-CAT as NODE-CAT-INS (presented in Algorithm~\ref{algo:nodecatins}) and NODE-CAT-COND (presented in Algorithm~\ref{algo:nodecatcond}), respectively.

The purpose of NODE-CAT-INS is to traverse the untraversed nodes, mark them as traversed, categorize them as broker nodes or community nodes and finally, place the broker nodes in brokerStack and community nodes in communityQueue. Categorization and placement of nodes is performed in a few steps. In line 2 of the procedure NODE-CAT-INS, we use a procedure called SPREAD, which spreads the influence from a node $v$ to all its uncovered neighbors.
NODE-CAT very often makes use of a procedure that calculates the $INS$ values of the neighbors of a node $v$, as described in Algorithm~\ref{algo:ins}. Similarly, NODE-CAT-COND, is also used for categorization of nodes. Note that we do not use SPREAD in the COND-based method.
In NODE-CAT-COND procedure, COND($u$) represents the subroutine to find out the value of conductance, as described in Algorithm~\ref{algo:cond}. Please note that for a network with multiple connected components, just like any other traversal method, LINCOM can also be applied sequentially to all the components.

Algorithm~\ref{algo:postprocess} describes the procedure POST-PROCESS, which places the broker nodes in the clusters of the present cover $G_s$. Community label of a broker node $v$ is assigned to the community that contributes to the largest fraction of community nodes in its neighborhood. If there is a tie or all the neighbors of a broker node consist only of brokers then the community label of the broker remains unassigned. Such brokers are assigned a community label during one of the latter procedure calls, referred to as MOD-MAXIMIZE. MOD-MAXIMIZE merges such brokers with the existing clusters such that the merged clusters would generate a cover with improved modularity. In MOD-MAXIMIZE, first we reduce the network by converting each of the initial clusters into a super-vertex. This is done by the procedure REDUCE. After reduction of the network, MOD-MAXIMIZE is applied in a way similar to the Louvain method~\citep{BGLL08}.


\SetAlFnt{\small}

\DontPrintSemicolon
\begin{algorithm}[htbp!] 
	\SetKwInOut{Input}{Input}
	\SetKwInOut{Output}{Output}
	\SetCommentSty{textrm}
	\SetKwComment{Comment}{$\triangleright$}{}
	\Input{ Undirected, unwtd. graph $G(V, E)$, threshold $(r)$}
	\Output{ Cover of $k$ communities, $G_s=\{G_{s_1}, G_{s_2},...,G_{s_k}\}$}
	\hrule
	\BlankLine
	\Begin
	{
		find $v_{start}$ $\in$ $V$, $\ni$ $v_{start}$ is the node with lowest degree\;
		\ForAll{$v \in V$}
		{
			$v.$nodeType $\leftarrow$  $v.$covered $\leftarrow$ 0\;
			$v.$community $\leftarrow$ $v$\;
		}
		$S$ $\leftarrow$ $Q$ $\leftarrow$ $\emptyset$ \Comment*{brokerStack($S$), communityQueue($Q$)}{}
		coverCount $\leftarrow$ 1\;
		$v$ $\leftarrow$ $v_{start}$\;
		NODE-CAT($G,v,Q,S$)\Comment*{\scriptsize{NODE-CAT-INS or NODE-CAT-COND}}{}
		\While{coverCount $<$ n}
		{
			\eIf{Q is non-empty}{
				$v$ $\leftarrow$ dequeue($Q$)\;
			}
			{
				$v$ $\leftarrow$ pop($S$)\;
			}
			NODE-CAT($G,v,Q,S$)\Comment*{\scriptsize{INS or COND-based NODE-CAT}}{}
		}
		\ForAll{$v \in V$}{check $v.$community to form $G_s$\;}
		POST-PROCESS($G_s$)\;
		REDUCE($G_s$)\;
		MOD-MAXIMIZE($G_s$)\;
		return $G_s$\;
	}
	\caption{Traversal-based Linear Time Community Detection (LINCOM($G,r$))\label{algo:lcd}}
\end{algorithm}

\DontPrintSemicolon
\begin{algorithm}[htbp!] 
	\SetKwInOut{Input}{Input}
	\SetKwInOut{Output}{Output}
	\SetCommentSty{textrm}
	\SetKwComment{Comment}{$\triangleright$}{}
	\Input{ Undirected, unwtd. graph $G(V, E)$, threshold $(r)$,
		\\\hspace{1mm} brokerStack $(S)$, communityQueue $(Q)$ }
	\Output{ Update $Q$, $S$}
	\hrule
	\BlankLine
	\Begin
	{
		SPREAD($v$)\;
		\For{$u \in \Gamma(v)$}
		{
			\lIf{$u.$nodeType $\ne 0$}
			{
				continue
			}
		
			\If{$\mathrm{INS}$$\mathrm{(}u\mathrm{)}$ $<$ $r$}
			{
				$u.$nodeType $\leftarrow$ 1 \Comment*{\footnotesize{marking the broker nodes}}{}
				push($S$,$u$)\;
			}
		
			\Else
			{
				$u.$nodeType $\leftarrow$ 2  \Comment*{\footnotesize{marking the community nodes}}{}
				enqueue($Q$,$u$)\;
				$u.$community $\leftarrow$ $v.$community\;

			}
		}
	}
	\caption{Categorizing uncovered nodes in $\Gamma(v)$ (NODE-CAT-INS($G,v,Q,S$))\label{algo:nodecatins}}
\end{algorithm}


\DontPrintSemicolon
\begin{algorithm}[htbp!] 
	\SetKwInOut{Input}{Input}
	\SetKwInOut{Output}{Output}
	\SetCommentSty{textrm}
	\SetKwComment{Comment}{$\triangleright$}{}
	\Input{ Undirected, unwtd. graph $G(V, E)$, root node $(v)$}
	\Output{ Update coverCount, covered list}
	\hrule
	\BlankLine
	\Begin
	{
		\For{$u \in \Gamma(v)$}
		{
			\If{$u.$covered $= 0$}
			{
				$u.$covered $\leftarrow$ 1\;
				coverCount $\leftarrow$ coverCount + 1\;
			}
		}
	}
	\caption{Spreading Influence to the Neighbors (SPREAD($v$))\label{algo:bfs}}
\end{algorithm}


\DontPrintSemicolon
\begin{algorithm}[htbp!] 
	\SetKwInOut{Input}{Input}
	\SetKwInOut{Output}{Output}
	\SetCommentSty{textrm}
	\SetKwComment{Comment}{$\triangleright$}{}
	\Input{ Undirected, unwtd. graph $G(V, E)$, root node $(v)$}
	\Output{ INS($v$)}
	\hrule
	\BlankLine
	\Begin
	{
		coveredCount $\leftarrow$ 0  \Comment*{\footnotesize{keeps a count of the covered neighbors of $v$}}{}
		\For{$u \in \Gamma(v)$}
		{
			\If{$u.$covered  $= 1$}
			{
				coveredCount $\leftarrow$ coveredCount + 1\;
			}
		}
		INS($v$) $\leftarrow$ coveredCount / deg($v$)\;
		return INS($v$)\;
	}
	\caption{Influenced Neighbors Score (INS($v$))\label{algo:ins}}
\end{algorithm}

\DontPrintSemicolon
\begin{algorithm}[htbp!] 
	\SetKwInOut{Input}{Input}
	\SetKwInOut{Output}{Output}
	\SetCommentSty{textrm}
	\SetKwComment{Comment}{$\triangleright$}{}
	\Input{Undirected, unwtd. graph $G(V, E)$, set of nodes $V_s$ in cluster $s$}
	\Output{Conductance score of the set of nodes $V_s$ in cluster $s$}
	\hrule
	\BlankLine
	\Begin
	{
		cutSize $\gets 0$ \;
		\ForAll{$u \in V_s$}
		{
			\For{$v \in \Gamma(u)$}
			{
				\If{$v \in V \setminus V_s$}
				{
					cutSize $\gets$ cutSize $+ 1$ \;
				}
			}
		}
		
		\Return{cutSize / min\{$\sum_{w \in V_s}^{} deg(w)$, $\sum_{w \in V \setminus V_s} deg(w)$\}}
	}
\caption{Conductance ($COND(s)$)}
\label{algo:cond}
\end{algorithm}

\DontPrintSemicolon
\begin{algorithm}[htbp!] 
	\SetKwInOut{Input}{Input}
	\SetKwInOut{Output}{Output}
	\SetCommentSty{textrm}
	\SetKwComment{Comment}{$\triangleright$}{}
	\Input{ Undirected, unwtd. graph $G(V, E)$, root node $(v)$}
	\Output{ Update $v.$community value for broker nodes}
	\hrule
	\BlankLine
	\Begin
	{
		\ForAll{$v \in V$}{
			$max(v)$ $\leftarrow$ 0\;
		}
		\ForAll{$v \in V$}{
			\If{$v.$nodeType $= 1$}{
				\ForAll{$G_{s_i}$ $\in$ $G_s$}{
					\eIf{$\frac{|\textit{Neighbors of v in }G_{s_i}|}{|G_{s_i}|}$ $>$ $max(v)$}{
						$max(v)$ $\leftarrow$ $\frac{|\textit{Neighbors of v in }G_{s_i}|}{|G_{s_i}|}$\;
						$v.$community $\leftarrow$ community label that leads to max($v$)\;
					}
					{
						\If{$\frac{|\textit{Neighbors of v in }G_{s_i}|}{|G_{s_i}|} = max(v)$}
						{
							$v.$community $\leftarrow$ append community label to community list\;
						}
					}
				}
			}
		}
	}
	\caption{Post-processing of the Broker Nodes (POST-PROCESS($G_s$))\label{algo:postprocess}}
\end{algorithm}

\DontPrintSemicolon
\begin{algorithm}[htbp!] 
	\SetKwInOut{Input}{Input}
	\SetKwInOut{Output}{Output}
	\SetCommentSty{textrm}
	\SetKwComment{Comment}{$\triangleright$}{}
	\Input{ $G_s$}
	\Output{ Updated $G_s$ ($G_s'$)}
	\hrule
	\BlankLine
	\Begin
	{
		\ForAll{$G_{s_i}$ $\in$ $G_s$}{
			$v_i'$ represents all the nodes in $V_{s_i}$\;
			$w(v_i', v_i')$ $\leftarrow$ $\Sigma_{u, v \in V_{s_i}} w(u, v)$\;
			$w(v_i', v_j')$ $\leftarrow$ $\Sigma_{u \in V_{s_i}, v \in V_{s_j}} w(u, v) | u \neq v, i \neq j$\;
		}
		return $G_s'$\;
	}
	\caption{Reduction of the Network (REDUCE($G_s$))\label{algo:reduce}}
\end{algorithm}

\DontPrintSemicolon
\begin{algorithm}[htbp!]  
	\SetKwInOut{Input}{Input}
	\SetKwInOut{Output}{Output}
	\SetCommentSty{textrm}
	\SetKwComment{Comment}{$\triangleright$}{}
	\Input{ $G_s$}
	\Output{ Updated $G_s$ ($G_s'$)}
	\hrule
	\BlankLine
	\Begin
	{
		\Repeat{there is no $\Delta Q_{max}$ $>$ 0}{
			\ForAll{$v$ $\in$ $G_s$}{
				\ForAll{$u$ $\in$ $\Gamma(v)$}{
					find $\Delta Q$ if $v$ moved to $u.$community \; 
				}
				\If{$\Delta Q_{max}$ $>$ 0}{
					move $v$ to $u_{max}.$community with $\Delta Q_{max}$\;
				}
			}
		}
		return $G_s'$\;
	}
	\caption{Modularity Maximization (MOD-MAXIMIZE($G_s$))\label{algo:modmax}}
\end{algorithm}

\DontPrintSemicolon
\begin{algorithm}[htbp!]  
	\SetKwInOut{Input}{Input}
	\SetKwInOut{Output}{Output}
	\SetCommentSty{textrm}
	\SetKwComment{Comment}{$\triangleright$}{}
	\Input{ Undirected, unwtd. graph $G(V, E)$, brokerStack $(S)$, communityQueue $(Q)$ }
	\Output{ Update $Q$, $S$}
	\hrule
	\BlankLine
	\Begin
	{
	    \For{$u \in \Gamma(v)$}
	    {
	        \lIf{$u.nodeType \ne 0$}
	        {continue}
	        \If{$u$ satisfies condition from Equation~\ref{eqn:cond}}
	        {
	            $u.nodeType \gets 2$  \Comment*{\footnotesize{marking the community nodes}}{}
	            $enqueue(Q,u)$ \;
	            $u.community \gets v.community$ \;
	        }
	        \Else
	        {
	            $u.nodeType \gets 1$ \Comment*{\footnotesize{marking the broker nodes}}{}
	            $push(S, u)$ \;
	        }
	    }
	}
	\caption{Categorizing uncovered nodes in $\Gamma(v)$ (NODE-CAT-COND($G,v,Q,S$))\label{algo:nodecatcond}}
\end{algorithm}

\textit{Complexity Analysis}: Our algorithm uses breadth first and depth first traversal techniques, both known to have worst case time complexities of O($|V|$ + $|E|$), where $|E|$ denotes the number of edges in the graph. The categorization of nodes takes O($|E|$) time as it involves traversal through all the edges. The time taken to place the broker depends upon the number of brokers obtained and their respective degrees. Number of brokers can never exceed the number of nodes and as the sum of degrees of all nodes is bounded by $2|E|$, the time complexity for this step is O($|E|$).\\
Lastly, the reduction procedure merges all the nodes in a community into one super-vertex and adds up all the inter-cluster and intra-cluster edges separately. Given, that community membership of a node can be accessed in constant running time, it has to traverse all the edges once. Therefore, the network reduction runs in O($|E|$). In modularity maximization part, we already work on a reduced graph and so its running time will not exceed the order of the reduced graph, because every node will need to traverse its neighbors trying to find the maximum modularity gain. Therefore, the overall worst case running time of our algorithm to create the initial bias for the community detection will be O($|E|$). The time complexity of the modularity maximization part is unknown~\citep{BGLL08} but as it works on a reduced graph it works very fast in practice. We proceed to verify the performance of our method by showing some of the experimental results that we performed on some benchmark datasets.

\section{An Example to Illustrate the Split Produced by INS-based Method}

\begin{figure*}[htb]
    \vspace*{0.35cm}
    \centering
    \subcaptionbox{Broker Stack(BS): \textit{N}, Community Queue(CQ): \textit{empty}. Starts from lowest degree node  $N$.\label{fig:slide3}}[0.3\textwidth]{\includegraphics[scale=0.22]{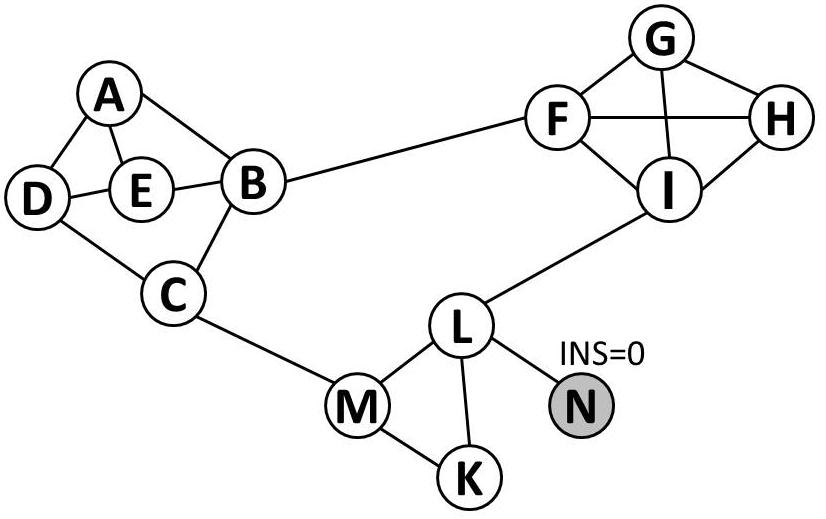}}
    \hfill
    \subcaptionbox{BS: \textit{L}, CQ:\textit{ empty}. Processing $N$. Neighbor of $N$ is $L$ (INS=0.25), goes to broker stack. \label{fig:slide4}} [0.3\textwidth]{\includegraphics[scale=0.22]{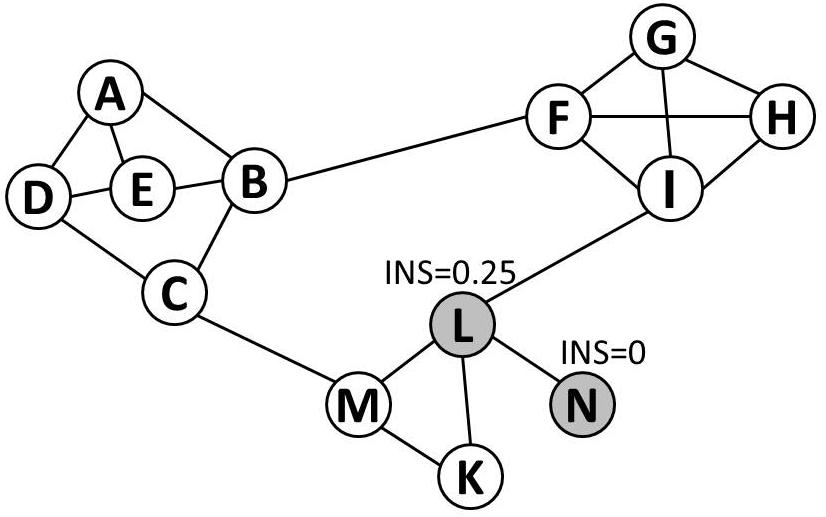}}
    \hfill
    \subcaptionbox{BS: \textit{I}, CQ: \textit{K  M}, Processing $L$. Neighbor of $M$ with INS $\geq$ 0.66,  goes to community queue. \label{fig:slide5}} [0.3\textwidth]{\includegraphics[scale=0.23]{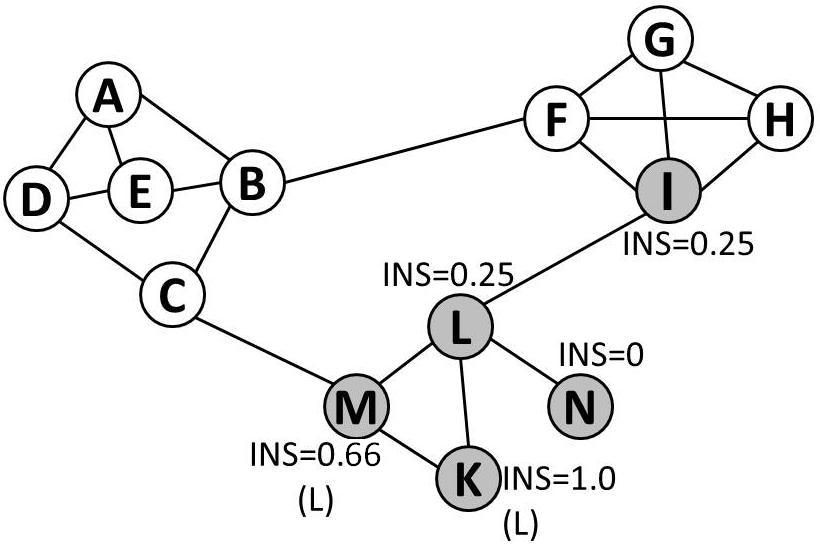}}
    \hfill
    
    \subcaptionbox{BS: \textit{I  C}, CQ: \textit{empty}. \label{fig:slide6}} [0.3\textwidth]{\includegraphics[scale=0.23]{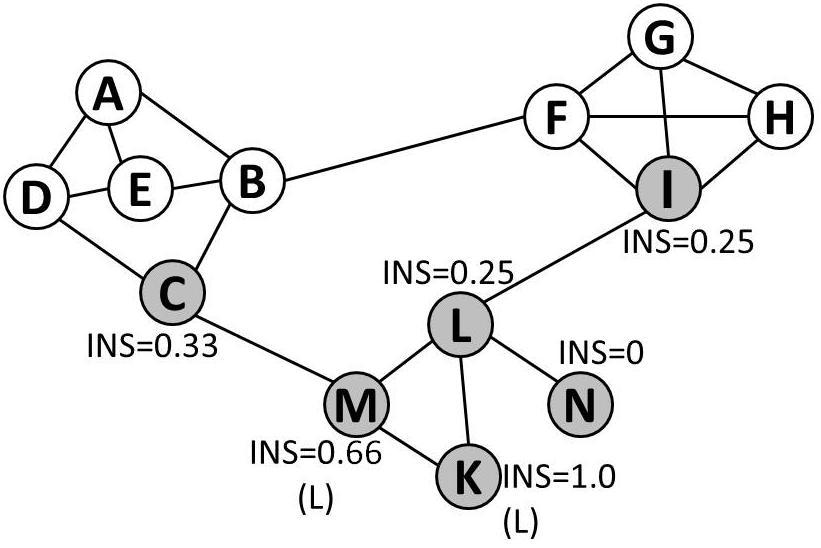}}
    \hfill
    \subcaptionbox{BS: \textit{I B D}, CQ: \textit{empty}, Processing $C$. \label{fig:slide7}} [0.3\textwidth]{\includegraphics[scale=0.23]{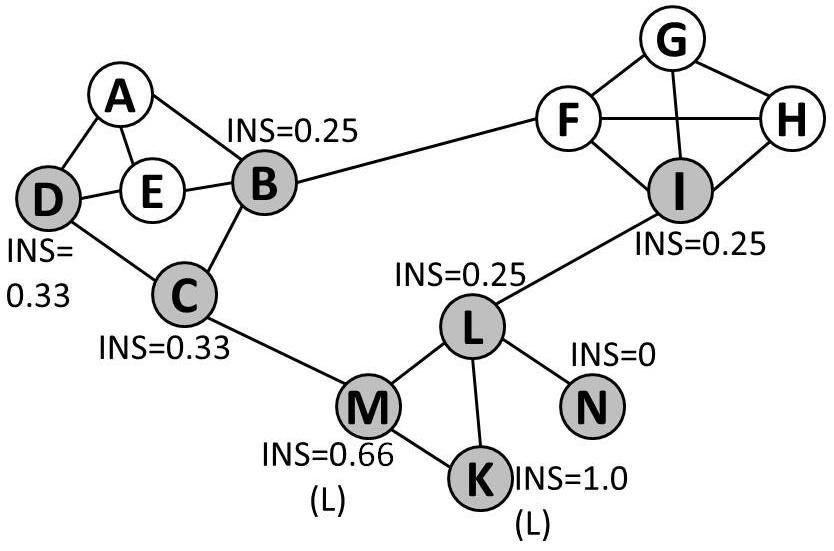}}
    \hfill
    \subcaptionbox{BS: \textit{I  B}, CQ: \textit{A  E}, Processing $D$. \label{fig:slide8}} [0.32\textwidth]{\includegraphics[scale=0.23]{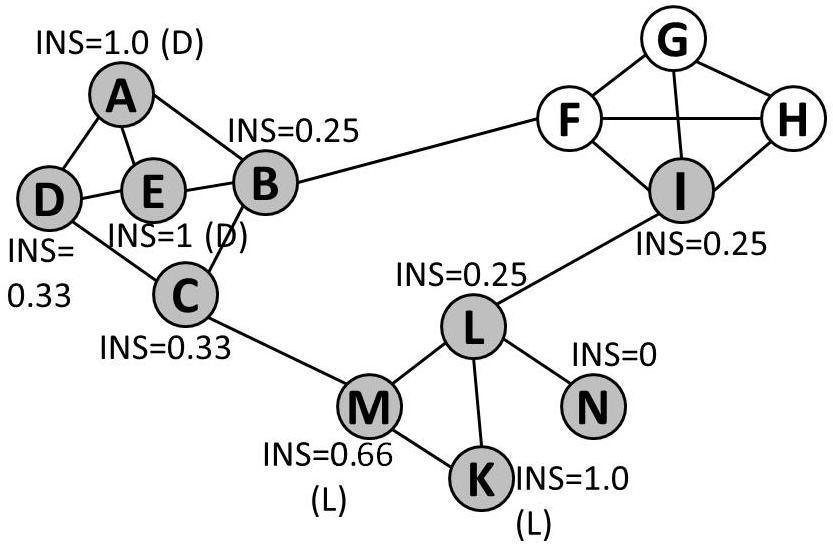}}
    \hfill
    
    \subcaptionbox{BS: \textit{I  F}, CQ: \textit{empty}, Processing $B$. \label{fig:slide9}} [0.32\textwidth]{\includegraphics[scale=0.23]{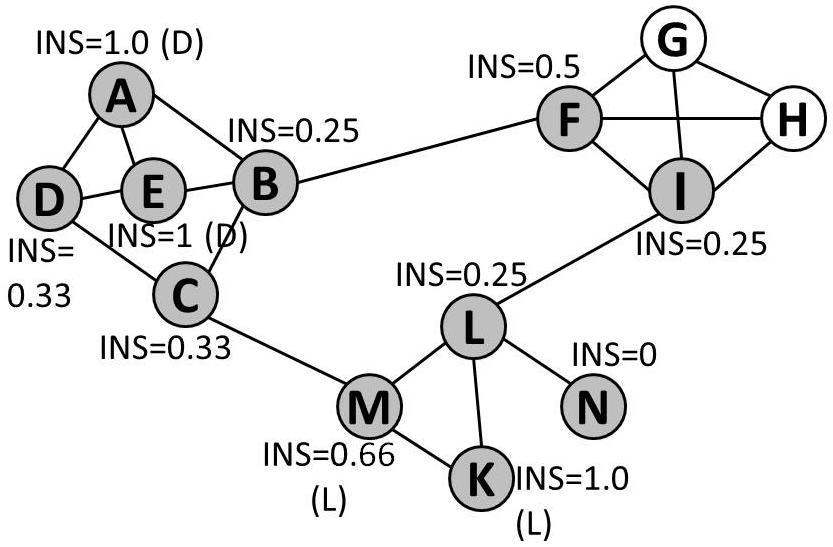}}
    \hfill
    \subcaptionbox{BS: \textit{I}, CQ: \textit{G H}, Processing $F$. \label{fig:slide10}} [0.3\textwidth]{\includegraphics[scale=0.23]{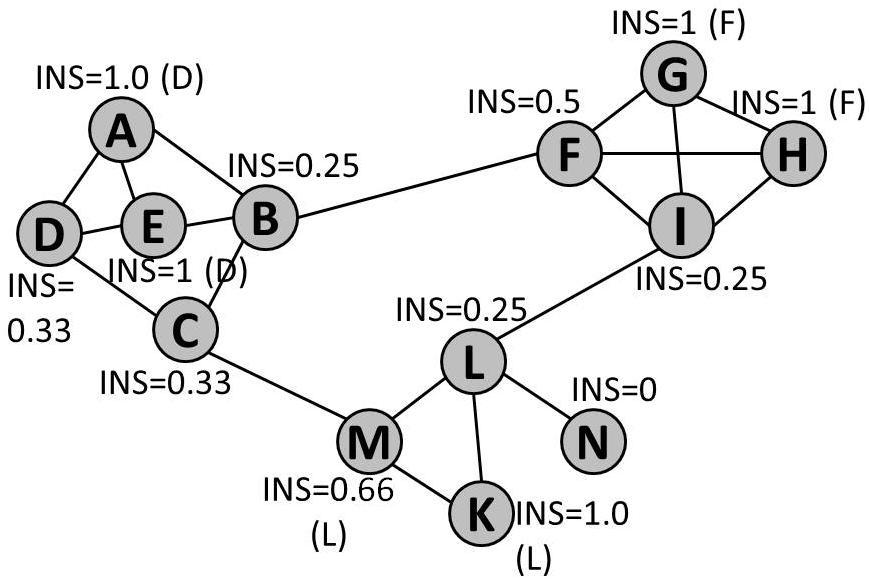}}
    \hfill
    \subcaptionbox{Three communities were found after modularity maximization. \label{fig:slide12}}[0.3\textwidth]{\includegraphics[scale=0.23]{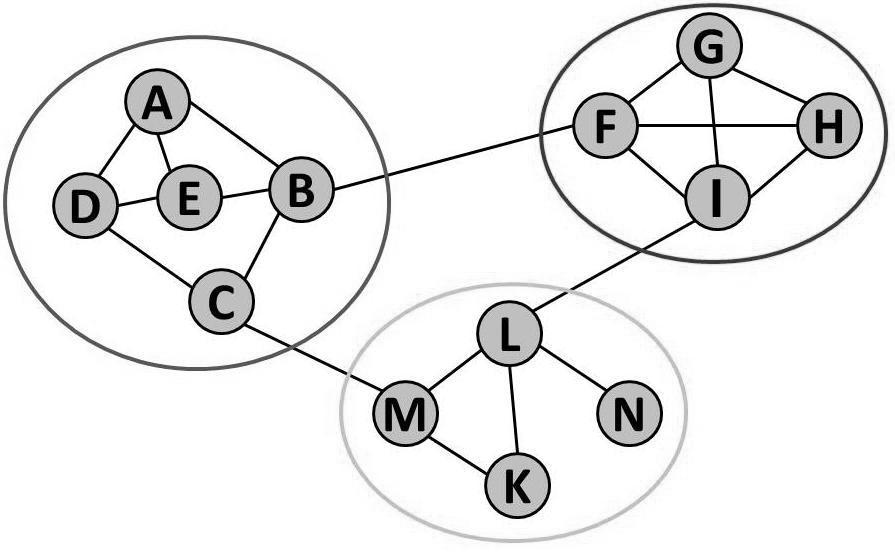}}
    \caption[Sequential steps to detect community using traversal-based algorithm.]{
    Steps to detect communities during the traversal-based algorithm LINCOM.
    }
    \label{fig:cd1}
\end{figure*}

We illustrate our proposed method with a graph consisting of 13 nodes and 20 edges. The nodes are labeled by uppercase letters, as shown in Figure~\ref{fig:slide3}. For this example, we take the threshold value of $r$ to be 0.66. As in OPTICS, our algorithm also starts from an outlier-like point, i.e., a pendant node, which does not play a significant role in a community. The starting node will always be considered as a broker node with an INS value of 0. Initially, from the only pendant node N, influence is propagated to L, as seen in Figures~\ref{fig:slide3} and ~\ref{fig:slide4}. INS(L) evaluates to 0.25 because only one node (N) out of all the four of its neighbors (I, K, M, N) has become influenced up to this stage. According to our algorithm, L is identified as a broker and pushed into the broker stack. Community queue is still empty at this stage. So we pop the top element from the broker stack, i.e., L, and process it. By processing a node $v$ we mean, propagating the influence to all the neighbors of $v$, then calculating INS value for all its neighbors, thereby categorizing them as broker nodes and community nodes and finally storing them in broker stack and community queue, respectively. For L, we spread the influence to I, K and M. INS values of I, K and M turn out to be 0.25, 1.0 and 0.67, respectively. Since INS(I) being less than the value of $r$ (0.66), I is placed in the broker stack and K and M are placed in community queue (we store them in lexicographic order, but it is not necessary). It should be noted that all the broker nodes discovered till this point have been assigned a community label that is same as their node label, as shown in Table~\ref{tab:example1}.

\begin{table}[ht]
\centering
\vspace{0.5cm}
\caption [Sequence of discovering the nodes from the given example]{\;Sequence of discovering the nodes in the example}
\ra{1.2}
\begin{tabular}{@{}lrlll@{}}
     \toprule
     Node & INS & Category & Initial Label & Final Label\\ \midrule 
     N & 0 & Broker & N & L \\ 
     L & 0.25 & Broker & L & L \\ 
     I & 0.25 & Broker & I & F \\ 
     K & 1 & Community Node & L & L \\ 
     M & 0.67 & Community Node & L & L \\ 
     C & 0.33 & Broker & C & D \\ 
     D & 0.33 & Broker & D & D \\ 
     B & 0.25 & Broker & B & D \\ 
     A & 1 & Community Node & D & D \\ 
     E & 1 & Community Node & D & D \\ 
     F & 0.50 & Broker & F & F \\ 
     G & 1 & Community Node & F & F \\ 
     H & 1 & Community Node & F & F \\
     \bottomrule
    \end{tabular}
    \label{tab:example1}
\end{table}

But when a community node is discovered, the node is given a community label same as the label of the last processed broker node, which is essentially the broker node that leads to its discovery. For example, here community(K) = community(M) = L. Now, after insertion of K and M, community queue is non-empty and hence the elements in the queue will be dequeued and processed until the queue is empty. The stack is not processed until the queue is empty. Therefore, K is processed first without spreading further influence to any other node in the graph. Then M reaches out to a single uncovered node C. INS(C) is 0.33 and it is subsequently placed in the stack. At this stage, the stack consists of I and C. Figures ~\ref{fig:slide5} and ~\ref{fig:slide6} show these steps. In Figure ~\ref{fig:slide7}, we see the broker node at the top of the stack is getting popped and thereby getting processed. C spreads influence to all its uncovered neighbors (B, D). Due to low INS values, B and D are both identified as broker nodes and get pushed into stack.
Now, after popping D and processing it, we reach nodes A and E. Both of them have the same INS value 1 and are identified as community nodes. The nodes in the graph's adjacency list are stored and processed in lexicographic order prompting B to be pushed to the stack before D. Therefore, D first comes to the top while popping and is processed before B. That is why A and E will also have a community label of D. This is shown in Figure~\ref{fig:slide8}. Next node to be processed is B. It spreads to a new node F. As INS(F) = 0.5, F is stored at the top of the stack. Community queue being empty, F is processed in the next step. F spreads to the remaining two uncovered nodes (G and H) of the network. For both of them, INS value evaluates to 1.0 and hence they are categorized as community nodes with community label F, as shown in Figures~\ref{fig:slide9} and~\ref{fig:slide10}. After post processing, the communities turn out to be as shown in Figure~\ref{fig:slide12}. Please note that this is an intermediate cover and not the final cover. The final cover is obtained by applying the procedure MOD-MAXIMIZE on the intermediate cover.

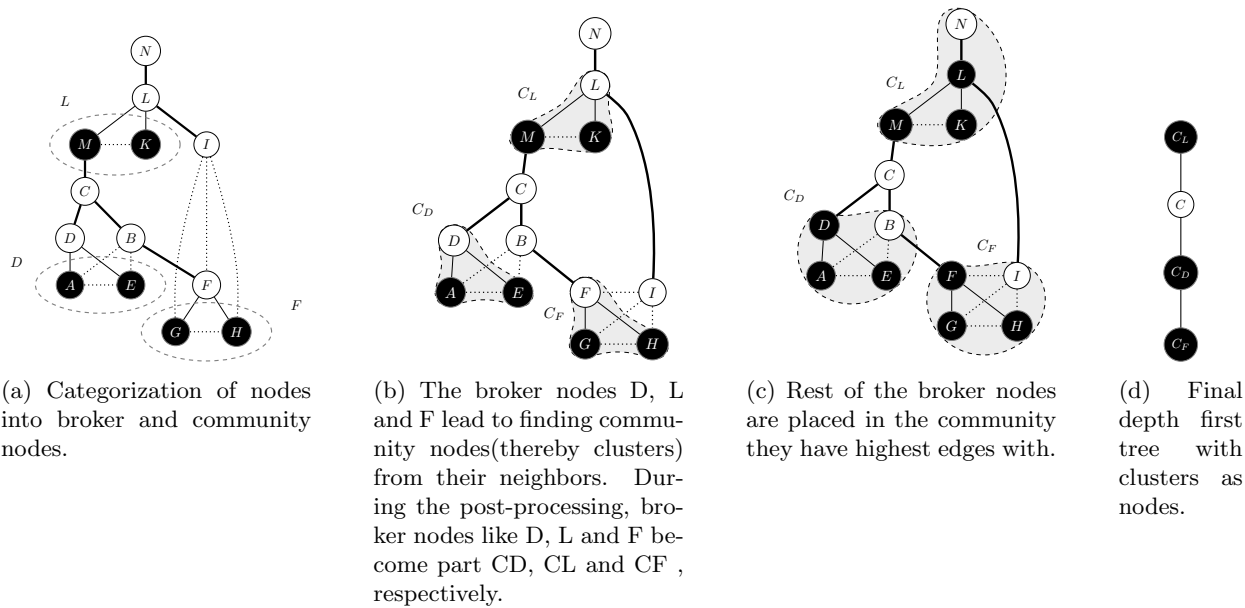
\begin{figure}[htb]
    \centering
    \vspace{0.35cm}
    \subcaptionbox{Categorization of nodes into broker and community nodes. \label{fig:stage1}}[0.25\textwidth]
    {\resizebox{0.25\textwidth}{!}
{
\begin{tikzpicture}[>=latex',line join=bevel,every fit/.style={ellipse,thick, dashed, draw,inner sep=4pt,text width=2cm}, scale=0.7]
\begin{scope}
\pgfsetstrokecolor{black}
\definecolor{strokecol}{rgb}{1.0,1.0,1.0};
\pgfsetstrokecolor{strokecol}
\definecolor{fillcol}{rgb}{1.0,1.0,1.0};
\pgfsetfillcolor{fillcol}
\filldraw (0.0bp,0.0bp) -- (0.0bp,318.0bp) -- (221.0bp,318.0bp) -- (221.0bp,0.0bp) -- cycle;
\end{scope}
\begin{scope}
\pgfsetstrokecolor{black}
\definecolor{strokecol}{rgb}{1.0,1.0,1.0};
\pgfsetstrokecolor{strokecol}
\definecolor{fillcol}{rgb}{1.0,1.0,1.0};
\pgfsetfillcolor{fillcol}
\filldraw (0.0bp,0.0bp) -- (0.0bp,318.0bp) -- (221.0bp,318.0bp) -- (221.0bp,0.0bp) -- cycle;
\end{scope}
\begin{scope}
\pgfsetstrokecolor{black}
\definecolor{strokecol}{rgb}{1.0,1.0,1.0};
\pgfsetstrokecolor{strokecol}
\definecolor{fillcol}{rgb}{1.0,1.0,1.0};
\pgfsetfillcolor{fillcol}
\filldraw (0.0bp,0.0bp) -- (0.0bp,318.0bp) -- (221.0bp,318.0bp) -- (221.0bp,0.0bp) -- cycle;
\end{scope}
\begin{scope}
\pgfsetstrokecolor{black}
\definecolor{strokecol}{rgb}{1.0,1.0,1.0};
\pgfsetstrokecolor{strokecol}
\definecolor{fillcol}{rgb}{1.0,1.0,1.0};
\pgfsetfillcolor{fillcol}
\filldraw (0.0bp,0.0bp) -- (0.0bp,318.0bp) -- (221.0bp,318.0bp) -- (221.0bp,0.0bp) -- cycle;
\end{scope}

\node (A) at (27.0bp,65.0bp) [draw=gray,fill=black,text=white!40,circle] {$A$};
\node (C) at (42.0bp,159.0bp) [draw=black,fill=white,circle] {$C$};
\node (B) at (88.0bp,112.0bp) [draw=black,fill=white,circle] {$B$};
\node (E) at (88.0bp,65.0bp) [draw=gray,fill=black,text=white!40,circle] {$E$};
\node (D) at (27.0bp,112.0bp) [draw=black,fill=white,circle] {$D$};
\node (G) at (133.0bp,18.0bp) [draw=gray,fill=black,text=white!40,circle] {$G$};
\node (F) at (164.0bp,65.0bp) [draw=black,fill=white,circle] {$F$};
\node (I) at (164.0bp,206.0bp) [draw=black,fill=white,circle] {$I$};
\node (H) at (194.0bp,18.0bp) [draw=gray,fill=black,text=white!40,circle] {$H$};
\node (K) at (103.0bp,206.0bp) [draw=gray,fill=black,text=white!40,circle] {$K$};
\node (M) at (42.0bp,206.0bp) [draw=gray,fill=black,text=white!40,circle] {$M$};
\node (L) at (103.0bp,253.0bp) [draw=black,fill=white,circle] {$L$};
\node (N) at (103.0bp,300.0bp) [draw=black,fill=white,circle] {$N$};

\draw [thick, dotted] (G) ..controls (162.37bp,18.0bp) and (164.57bp,18.0bp)  .. (H);
\draw [black,ultra thick] (M) ..controls (42.0bp,184.4bp) and (42.0bp,180.62bp)  .. (C);
\draw [solid] (L) ..controls (77.193bp,232.96bp) and (67.526bp,225.83bp)  .. (M);
\draw [black,ultra thick] (N) ..controls (103.0bp,278.4bp) and (103.0bp,274.62bp)  .. (L);
\draw [thick, dotted] (I) ..controls (164.0bp,161.28bp) and (164.0bp,109.95bp)  .. (F);
\draw [thick, dotted] (B) ..controls (62.193bp,91.962bp) and (52.526bp,84.83bp)  .. (A);
\draw [thick, dotted] (B) ..controls (88.0bp,90.403bp) and (88.0bp,86.622bp)  .. (E);
\draw [] (F) ..controls (150.15bp,43.902bp) and (146.89bp,39.169bp)  .. (G);
\draw [thick, dotted] (A) ..controls (56.367bp,65.0bp) and (58.57bp,65.0bp)  .. (E);
\draw [black,ultra thick] (B) ..controls (118.92bp,92.693bp) and (133.16bp,84.261bp)  .. (F);
\draw [] (D) ..controls (52.807bp,91.962bp) and (62.474bp,84.83bp)  .. (E);
\draw [] (D) ..controls (27.0bp,90.403bp) and (27.0bp,86.622bp)  .. (A);
\draw [thick, dotted] (I) ..controls (176.79bp,164.98bp) and (189.36bp,121.33bp)  .. (194.0bp,83.0bp) .. controls (195.91bp,67.226bp) and (195.69bp,49.105bp)  .. (H);
\draw [black,ultra thick] (C) ..controls (61.924bp,138.51bp) and (68.007bp,132.56bp)  .. (B);
\draw [black,ultra thick] (L) ..controls (128.81bp,232.96bp) and (138.47bp,225.83bp)  .. (I);
\draw [thick, dotted] (I) ..controls (151.39bp,164.95bp) and (138.94bp,121.29bp)  .. (134.0bp,83.0bp) .. controls (131.97bp,67.24bp) and (131.89bp,49.117bp)  .. (G);
\draw [thick, dotted] (M) ..controls (71.367bp,206.0bp) and (73.57bp,206.0bp)  .. (K);
\draw [black,ultra thick] (C) ..controls (35.148bp,137.45bp) and (33.804bp,133.41bp)  .. (D);
\draw [] (L) ..controls (103.0bp,231.4bp) and (103.0bp,227.62bp)  .. (K);
\draw [] (F) ..controls (177.4bp,43.902bp) and (180.55bp,39.169bp)  .. (H);

\node [gray,fit=(M) (K), label={[label distance=0.25cm]140:$L$}] {};
\node [gray,fit=(A) (E), label={[label distance=0.3cm]170:$D$}] {};
\node [gray,fit=(G) (H), label={[label distance=-4cm]200:$F$},] {};

\end{tikzpicture}		
}}
    \hfill
    \subcaptionbox{The broker nodes D, L and F lead to finding community nodes(thereby  clusters) from their neighbors. During the post-processing, broker nodes like D, L and F become part CD, CL and CF , respectively. \label{fig:stage2}}[0.25\textwidth]
    {\resizebox{0.25\textwidth}{!}
{
\begin{tikzpicture}[>=latex',line join=bevel,every fit/.style={regular polygon, regular polygon sides=3,thick, dashed, draw,inner sep=4pt,text width=2cm}, scale=0.72]
	\begin{scope}
	\pgfsetstrokecolor{black}
	\definecolor{strokecol}{rgb}{1.0,1.0,1.0};
	\pgfsetstrokecolor{strokecol}
	\definecolor{fillcol}{rgb}{1.0,1.0,1.0};
	\pgfsetfillcolor{fillcol}
	\end{scope}
	\begin{scope}
	\pgfsetstrokecolor{black}
	\definecolor{strokecol}{rgb}{1.0,1.0,1.0};
	\pgfsetstrokecolor{strokecol}
	\definecolor{fillcol}{rgb}{1.0,1.0,1.0};
	\pgfsetfillcolor{fillcol}
	\end{scope}
	\begin{scope}
	\pgfsetstrokecolor{black}
	\definecolor{strokecol}{rgb}{1.0,1.0,1.0};
	\pgfsetstrokecolor{strokecol}
	\definecolor{fillcol}{rgb}{1.0,1.0,1.0};
	\pgfsetfillcolor{fillcol}
	\end{scope}
	\begin{scope}
	\pgfsetstrokecolor{black}
	\definecolor{strokecol}{rgb}{1.0,1.0,1.0};
	\pgfsetstrokecolor{strokecol}
	\definecolor{fillcol}{rgb}{1.0,1.0,1.0};
	\pgfsetfillcolor{fillcol}
	\end{scope}
	\node (A) at (95.547bp,65.0bp) [draw=gray,fill=black,text=white!40,circle] {$A$};
	\node (C) at (159.55bp,159.0bp) [draw=black,fill=white,circle] {$C$};
	\node (B) at (159.55bp,112.0bp) [draw=black,fill=white,circle] {$B$};
	\node (E) at (156.55bp,65.0bp) [draw=gray,fill=black,text=white!40,circle] {$E$};
	\node (D) at (98.547bp,112.0bp) [draw=black,fill=white,circle] {$D$};
	\node (G) at (217.55bp,18.0bp) [draw=gray,fill=black,text=white!40,circle] {$G$};
	\node (F) at (217.55bp,65.0bp) [draw=black,fill=white,circle] {$F$};
	\node (I) at (278.55bp,65.0bp) [draw=black,fill=white,circle] {$I$};
	\coordinate (rank2) at (30.547bp,65.0bp);
	\node (K) at (226.55bp,206.0bp) [draw=gray,fill=black,text=white!40,circle] {$K$};
	\node (M) at (165.55bp,206.0bp) [draw=gray,fill=black,text=white!40,circle] {$M$};
	\node (L) at (226.55bp,253.0bp) [draw=black,fill=white,circle] {$L$};
	\node (N) at (226.55bp,300.0bp) [draw=black,fill=white,circle] {$N$};
	\node (H) at (278.55bp,18.0bp) [draw=gray,fill=black,text=white!40,circle] {$H$};
	\coordinate (rank1) at (30.547bp,112.0bp);
	\draw [white,] (rank2) ..controls (63.68bp,65.0bp) and (66.09bp,65.0bp)  .. (A);
	\draw [black, ultra thick] (N) ..controls (226.55bp,278.4bp) and (226.55bp,274.62bp)  .. (L);
	\draw [thick, dotted] (B) ..controls (132.99bp,92.328bp) and (122.5bp,84.95bp)  .. (A);
	\draw [white,] (E) ..controls (185.91bp,65.0bp) and (188.12bp,65.0bp)  .. (F);
	\draw [thick, dotted] (I) ..controls (278.55bp,43.403bp) and (278.55bp,39.622bp)  .. (H);
	\draw [black, ultra thick] (L) ..controls (249.01bp,234.84bp) and (253.59bp,229.69bp)  .. (256.55bp,224.0bp) .. controls (280.68bp,177.55bp) and (281.17bp,113.57bp)  .. (I);
	\draw [black, ultra thick] (B) ..controls (184.12bp,91.935bp) and (192.89bp,85.133bp)  .. (F);
	\draw [black, ultra thick] (C) ..controls (133.74bp,138.96bp) and (124.07bp,131.83bp)  .. (D);
	\draw [black, ultra thick] (M) ..controls (162.8bp,184.4bp) and (162.3bp,180.62bp)  .. (C);
	\draw [solid] (L) ..controls (200.74bp,232.96bp) and (191.07bp,225.83bp)  .. (M);
	\draw [] (F) ..controls (217.55bp,43.403bp) and (217.55bp,39.622bp)  .. (G);
	\draw [] (D) ..controls (97.174bp,90.403bp) and (96.922bp,86.622bp)  .. (A);
	\draw [thick, dotted] (A) -- (E);
	\draw [thick, dotted] (M) ..controls (194.91bp,206.0bp) and (197.12bp,206.0bp)  .. (K);
	\draw [white,->] (rank1) ..controls (30.547bp,93.763bp) and (30.547bp,93.539bp)  .. (rank2);
	\draw [] (D) ..controls (123.12bp,91.935bp) and (131.89bp,85.133bp)  .. (E);
	\draw [black, ultra thick] (C) ..controls (159.55bp,137.4bp) and (159.55bp,133.62bp)  .. (B);
	\draw [thick, dotted] (I) -- (F);
	\draw [thick, dotted] (I) ..controls (252.74bp,44.962bp) and (243.07bp,37.83bp)  .. (G);
	\draw [] (L) ..controls (226.55bp,231.4bp) and (226.55bp,227.62bp)  .. (K);
	\draw [] (F) ..controls (243.35bp,44.962bp) and (253.02bp,37.83bp)  .. (H);
	\draw [thick, dotted] (B) ..controls (158.17bp,90.403bp) and (157.92bp,86.622bp)  .. (E);
	\draw [thick, dotted] (G) ..controls (246.91bp,18.0bp) and (249.12bp,18.0bp)  .. (H);
	
	\node[above of=M] {$C_L$};
	\node[above left of=D] {$C_D$};
	\node[above left of=G] {$C_F$};
	
	\begin{pgfonlayer}{background}
	\draw[black, fill=gray!15, dashed](L.north east) to[closed,curve through={($(L.north east)!0.5!(K.east)$) .. (K.east) ..(M.south east) .. (M.south west) .. ($(M.north east)!0.5!(L.west)$) }] (L.north);
	
	\draw[black, fill=gray!15, dashed](D.north west) to[closed,curve through={($(D.south west)!0.5!(A.north west)$) .. (A.south west) .. (A.south east) ..(E.south west) .. (E.south east) .. ($(E.north east)!0.5!(D.south east)$) }] (D.north);
	
	\draw[black, fill=gray!15, dashed](F.north west) to[closed,curve through={($(F.south west)!0.5!(G.north west)$) .. (G.south west) .. (G.south east) ..(H.south west) .. (H.south east) .. ($(H.north east)!0.5!(F.south east)$) }] (F.north);
	\end{pgfonlayer}
	%
\end{tikzpicture}
}}
    \hfill
    \subcaptionbox{Rest of the broker nodes are placed in the community they have highest edges with. \label{fig:stage3}}[0.25\textwidth]
    {\resizebox{0.25\textwidth}{!}
{
\begin{tikzpicture}[>=latex',line join=bevel,every fit/.style={regular polygon, regular polygon sides=3,thick, dashed, draw,inner sep=4pt,text width=2cm}, scale=0.72]
	\begin{scope}
	\pgfsetstrokecolor{black}
	\definecolor{strokecol}{rgb}{1.0,1.0,1.0};
	\pgfsetstrokecolor{strokecol}
	\definecolor{fillcol}{rgb}{1.0,1.0,1.0};
	\pgfsetfillcolor{fillcol}
	\end{scope}
	\begin{scope}
	\pgfsetstrokecolor{black}
	\definecolor{strokecol}{rgb}{1.0,1.0,1.0};
	\pgfsetstrokecolor{strokecol}
	\definecolor{fillcol}{rgb}{1.0,1.0,1.0};
	\pgfsetfillcolor{fillcol}
	\end{scope}
	\begin{scope}
	\pgfsetstrokecolor{black}
	\definecolor{strokecol}{rgb}{1.0,1.0,1.0};
	\pgfsetstrokecolor{strokecol}
	\definecolor{fillcol}{rgb}{1.0,1.0,1.0};
	\pgfsetfillcolor{fillcol}
	\end{scope}
	\begin{scope}
	\pgfsetstrokecolor{black}
	\definecolor{strokecol}{rgb}{1.0,1.0,1.0};
	\pgfsetstrokecolor{strokecol}
	\definecolor{fillcol}{rgb}{1.0,1.0,1.0};
	\pgfsetfillcolor{fillcol}
	\end{scope}
	\node (A) at (95.547bp,65.0bp) [draw=gray,fill=black,text=white,circle] {$A$};
	\node (C) at (159.55bp,159.0bp) [draw=black, fill=white,circle] {$C$};
	\node (B) at (159.55bp,112.0bp) [draw=black, fill=white,circle] {$B$};
	\node (E) at (156.55bp,65.0bp) [draw=gray,fill=black,text=white,circle] {$E$};
	\node (D) at (98.547bp,112.0bp) [draw=gray,fill=black,text=white,circle] {$D$};
	\node (G) at (217.55bp,18.0bp) [draw=gray,fill=black,text=white,circle] {$G$};
	\node (F) at (217.55bp,65.0bp) [draw=gray,fill=black,text=white,circle] {$F$};
	\node (I) at (278.55bp,65.0bp) [draw=black, fill=white,circle] {$I$};
	\coordinate (rank2) at (30.547bp,65.0bp);
	\node (K) at (226.55bp,206.0bp) [draw=gray,fill=black,text=white,circle] {$K$};
	\node (M) at (165.55bp,206.0bp) [draw=gray,fill=black,text=white,circle] {$M$};
	\node (L) at (226.55bp,253.0bp) [draw=gray,fill=black,text=white,circle] {$L$};
	\node (N) at (226.55bp,300.0bp) [draw=black, fill=white,circle] {$N$};
	\node (H) at (278.55bp,18.0bp) [draw=gray,fill=black,text=white,circle] {$H$};
	\coordinate (rank1) at (30.547bp,112.0bp);
	\draw [white,] (rank2) ..controls (63.68bp,65.0bp) and (66.09bp,65.0bp)  .. (A);
	\draw [black, ultra thick] (N) ..controls (226.55bp,278.4bp) and (226.55bp,274.62bp)  .. (L);
	\draw [thick, dotted] (B) ..controls (132.99bp,92.328bp) and (122.5bp,84.95bp)  .. (A);
	\draw [white,] (E) ..controls (185.91bp,65.0bp) and (188.12bp,65.0bp)  .. (F);
	\draw [thick, dotted] (I) ..controls (278.55bp,43.403bp) and (278.55bp,39.622bp)  .. (H);
	\draw [black, ultra thick] (L) ..controls (249.01bp,234.84bp) and (253.59bp,229.69bp)  .. (256.55bp,224.0bp) .. controls (280.68bp,177.55bp) and (281.17bp,113.57bp)  .. (I);
	\draw [black, ultra thick] (B) ..controls (184.12bp,91.935bp) and (192.89bp,85.133bp)  .. (F);
	\draw [black, ultra thick] (C) ..controls (133.74bp,138.96bp) and (124.07bp,131.83bp)  .. (D);
	\draw [black, ultra thick] (M) ..controls (162.8bp,184.4bp) and (162.3bp,180.62bp)  .. (C);
	\draw [solid] (L) ..controls (200.74bp,232.96bp) and (191.07bp,225.83bp)  .. (M);
	\draw [] (F) ..controls (217.55bp,43.403bp) and (217.55bp,39.622bp)  .. (G);
	\draw [] (D) ..controls (97.174bp,90.403bp) and (96.922bp,86.622bp)  .. (A);
	\draw [thick, dotted] (A) -- (E);
	\draw [thick, dotted] (M) ..controls (194.91bp,206.0bp) and (197.12bp,206.0bp)  .. (K);
	\draw [white,->] (rank1) ..controls (30.547bp,93.763bp) and (30.547bp,93.539bp)  .. (rank2);
	\draw [] (D) ..controls (123.12bp,91.935bp) and (131.89bp,85.133bp)  .. (E);
	\draw [black, ultra thick] (C) ..controls (159.55bp,137.4bp) and (159.55bp,133.62bp)  .. (B);
	\draw [thick, dotted] (I) -- (F);
	\draw [thick, dotted] (I) ..controls (252.74bp,44.962bp) and (243.07bp,37.83bp)  .. (G);
	\draw [] (L) ..controls (226.55bp,231.4bp) and (226.55bp,227.62bp)  .. (K);
	\draw [] (F) ..controls (243.35bp,44.962bp) and (253.02bp,37.83bp)  .. (H);
	\draw [thick, dotted] (B) ..controls (158.17bp,90.403bp) and (157.92bp,86.622bp)  .. (E);
	\draw [thick, dotted] (G) ..controls (246.91bp,18.0bp) and (249.12bp,18.0bp)  .. (H);
	
	\node[above of=M] {$C_L$};
	\node[above left of=D] {$C_D$};
	\node[above left of=I] {$C_F$};
	
	\begin{pgfonlayer}{background}
	\draw[black, fill=gray!15, dashed](N.north east) to[closed,curve through={ (K.south east) ..(M.south) .. (M.south west) .. ($(M.north)!0.7!(L.west)$)}] (N.north);
	
	\draw[black, fill=gray!15, dashed](D.north west) to[closed,curve through={(D.north east) .. (B.north east) .. (E.south east) .. (A.south west) .. (D.north west)}] (D.north west);
	
	\draw[black, fill=gray!15, dashed](F.north west) to[closed,curve through={(F.north east) .. (I.north east) .. (H.south east) .. (G.south west) .. (F.north west)}] (F.north west);
	\end{pgfonlayer}
	%
	\end{tikzpicture}
}}
    \hfill
    \subcaptionbox{Final depth first tree with clusters as nodes. \label{fig:stage4}}[0.10\textwidth]
    {\resizebox{0.035\textwidth}{!}
{
\begin{tikzpicture}[>=latex',line join=bevel, scale=0.2]
	\node[circle, draw=gray, fill=black, text=white] (C_L) {$C_L$};

	\node[circle, draw=black, fill=white, text=black, below=of C_L] (C) {$C$};
	
	\node[circle, draw=gray, fill=black, text=white, below=of C] (C_D) {$C_D$};
	
	\node[circle, draw=gray, fill=black, text=white, below=of C_D] (C_F) {$C_F$};
	
	\draw[thick, draw=black] (C_L) -- (C) -- (C_D) -- (C_F);
	
\end{tikzpicture}
}}
    
    \caption{Transforming the network into a depth first tree of clusters. Breadth first traversal is used when traversing within the communities. The white nodes are the broker nodes and the black nodes are the community nodes. The thick lines mark the discovery of a broker node.}
    \label{fig:dfsbfs}
\end{figure}



It is important to understand that the initial clustering generated by LINCOM is done by running two graph traversal methods, i.e., depth first traversal and breadth first traversal, in parallel. We start from a node assuming it is getting the spread started. Since it is getting the spread started, irrespective of the node actually being a broker node or a node inside a community, it will be considered as a broker node (once the traversals are over, during the post-processing, it will be placed into the correct community eventually). The traversal methods and the corresponding trees have been shown in Figure~\ref{fig:dfsbfs}. In Figure~\ref{fig:stage1}, we see how the nodes have been categorized into two different types - namely the broker nodes (white) and the community nodes (black). Note that the community nodes store the label of the broker node from which is was discovered. In Figure~\ref{fig:stage2}, such broker nodes, that are used to label a community, is placed inside that community. In Figure~\ref{fig:stage3}, the other broker nodes are placed inside the community, which has most number of edges connected to it. If there is a tie, then the broker node is considered as a singleton cluster and is passed on to modularity maximization process as a part of the initial cover. In Figure~\ref{fig:stage4}, we observe that node C has equal number edges to both $C_D$ and $C_L$.  Modularity maximization agglomerates such a node to a cluster such that the overall modularity of the final cover is maximized.

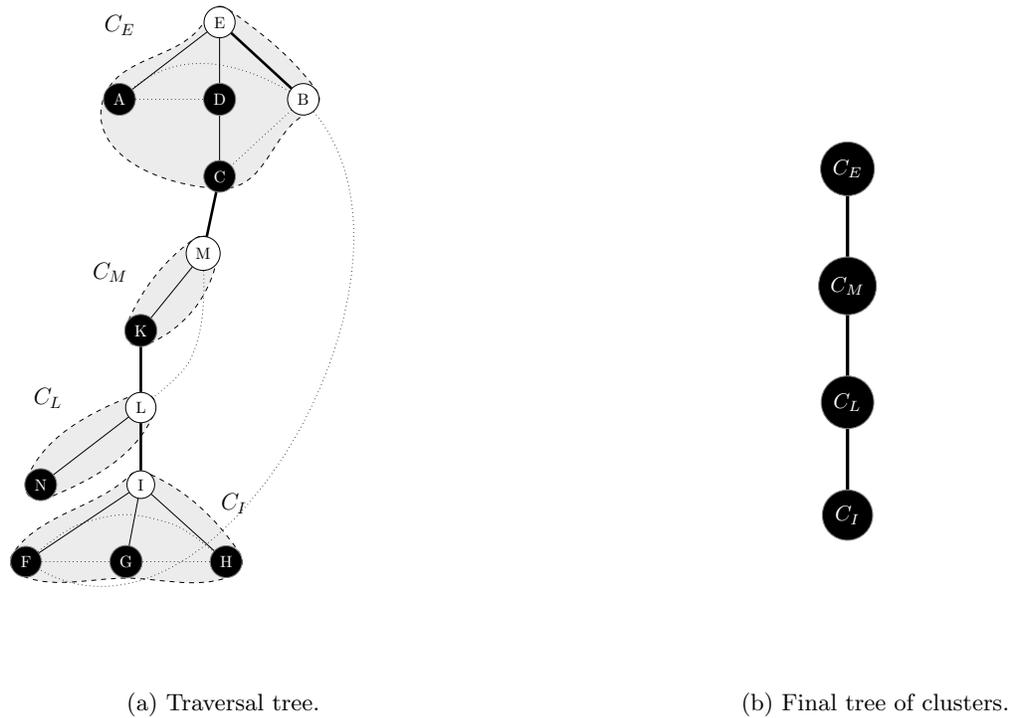
\begin{figure}[htb]
    \centering
    \vspace{0.75cm}
    \subcaptionbox{Traversal tree. \label{fig:stage5}}[0.70\textwidth]
    {\resizebox{0.37\textwidth}{!}
{
\raisebox{-1cm}{
\begin{tikzpicture}[>=latex',line join=bevel,every edge/.style={very thick}]
\begin{scope}
  \pgfsetstrokecolor{black}
  \definecolor{strokecol}{rgb}{1.0,1.0,1.0};
  \pgfsetstrokecolor{strokecol}
  \definecolor{fillcol}{rgb}{1.0,1.0,1.0};
  \pgfsetfillcolor{fillcol}
\end{scope}
\begin{scope}
  \pgfsetstrokecolor{black}
  \definecolor{strokecol}{rgb}{1.0,1.0,1.0};
  \pgfsetstrokecolor{strokecol}
  \definecolor{fillcol}{rgb}{1.0,1.0,1.0};
  \pgfsetfillcolor{fillcol}
\end{scope}
\begin{scope}
  \pgfsetstrokecolor{black}
  \definecolor{strokecol}{rgb}{1.0,1.0,1.0};
  \pgfsetstrokecolor{strokecol}
  \definecolor{fillcol}{rgb}{1.0,1.0,1.0};
  \pgfsetfillcolor{fillcol}
\end{scope}
\begin{scope}
  \pgfsetstrokecolor{black}
  \definecolor{strokecol}{rgb}{1.0,1.0,1.0};
  \pgfsetstrokecolor{strokecol}
  \definecolor{fillcol}{rgb}{1.0,1.0,1.0};
  \pgfsetfillcolor{fillcol}
\end{scope}
\begin{scope}
  \pgfsetstrokecolor{black}
  \definecolor{strokecol}{rgb}{1.0,1.0,1.0};
  \pgfsetstrokecolor{strokecol}
  \definecolor{fillcol}{rgb}{1.0,1.0,1.0};
  \pgfsetfillcolor{fillcol}
\end{scope}
\begin{scope}
  \pgfsetstrokecolor{black}
  \definecolor{strokecol}{rgb}{1.0,1.0,1.0};
  \pgfsetstrokecolor{strokecol}
  \definecolor{fillcol}{rgb}{1.0,1.0,1.0};
  \pgfsetfillcolor{fillcol}
\end{scope}
\begin{scope}
  \pgfsetstrokecolor{black}
  \definecolor{strokecol}{rgb}{1.0,1.0,1.0};
  \pgfsetstrokecolor{strokecol}
  \definecolor{fillcol}{rgb}{1.0,1.0,1.0};
  \pgfsetfillcolor{fillcol}
\end{scope}
  \node (A) at (98.297bp,347.0bp) [draw=gray,fill=black,text=white,circle] {A};
  \node (C) at (159.3bp,300.0bp) [draw=gray,fill=black,text=white,circle] {C};
    \node (D) at (159.3bp,347.0bp) [draw=gray,fill=black,text=white,circle] {D};
  \node (B) [draw=black,fill=white,circle, right of=D, xshift=0.8cm] {B};
  \node (E) at (159.3bp,394.0bp) [draw=black,fill=white,circle] {E};

  \node (G) at (102.3bp,65.0bp) [draw=gray,fill=black,text=white,circle] {G};
  \node (F) at (41.297bp,65.0bp) [draw=gray,fill=black,text=white,circle] {F};
  \node (I) at (111.3bp,112.0bp) [draw=black,fill=white,circle] {I};
  \node (H) at (163.3bp,65.0bp) [draw=gray,fill=black,text=white,circle] {H};
  \node (K) at (111.3bp,206.0bp) [draw=gray,fill=black,text=white,circle] {K};
  \node (M) at (149.3bp,253.0bp) [draw=black,fill=white,circle] {M};
  \node (L) at (111.3bp,159.0bp) [draw=black,fill=white,circle] {L};
  \node (N) at (50.297bp,112.0bp) [draw=gray,fill=black,text=white,circle] {N};

  \draw [] (I) ..controls (107.18bp,90.403bp) and (106.42bp,86.622bp)  .. (G);
  \draw [dotted] (F) ..controls (70.664bp,65.0bp) and (72.867bp,65.0bp)  .. (G);
  \draw [ultra thick, draw=black] (L) ..controls (111.3bp,137.4bp) and (111.3bp,133.62bp)  .. (I);
  \draw [] (E) ..controls (159.3bp,372.4bp) and (159.3bp,368.62bp)  .. (D);
  \draw [ultra thick, draw=black] (K) ..controls (111.3bp,184.4bp) and (111.3bp,180.62bp)  .. (L);
  \draw [] (L) ..controls (85.49bp,138.96bp) and (75.823bp,131.83bp)  .. (N);
  \draw [dotted] (G) ..controls (131.66bp,65.0bp) and (133.87bp,65.0bp)  .. (H);
  \draw [] (I) ..controls (133.63bp,91.671bp) and (141.07bp,85.236bp)  .. (H);
  \draw [dotted] (M) ..controls (150.29bp,221.33bp) and (148.73bp,202.56bp)  .. (141.3bp,188.0bp) .. controls (138.38bp,182.29bp) and (133.81bp,177.13bp)  .. (L);
  \draw [dotted] (A) ..controls (127.66bp,347.0bp) and (129.87bp,347.0bp)  .. (D);
  \draw [draw=black, ultra thick] (C) ..controls (154.72bp,278.4bp) and (153.88bp,274.62bp)  .. (M);
  \draw [] (M) ..controls (132.58bp,232.21bp) and (128.09bp,226.89bp)  .. (K);
  \draw [dotted] (F) ..controls (63.45bp,83.104bp) and (69.564bp,86.579bp)  .. (75.797bp,88.5bp) .. controls (98.308bp,95.438bp) and (106.29bp,95.438bp)  .. (128.8bp,88.5bp) .. controls (135.03bp,86.579bp) and (141.14bp,83.104bp)  .. (H);
  \draw [] (D) ..controls (159.3bp,325.4bp) and (159.3bp,321.62bp)  .. (C);
  \draw [] (I) ..controls (82.495bp,92.484bp) and (70.159bp,84.554bp)  .. (F);
  \draw [] (E) ..controls (133.49bp,373.96bp) and (123.82bp,366.83bp)  .. (A);

\draw[ultra thick, draw=black] (E) -- (B);
\draw[dotted] (B) to[out=-50,in=-30] (F);
\draw (B)[dotted] -- (C);
\draw (B)[dotted] to[in=40, out=150] (A);

	\node[draw=none, fill=none, above of= A, yshift=0.6cm](AA) {\Large $C_E$};
	\node[draw=none, fill=none, left of= M, yshift=-0.4cm, xshift=-1cm](MA) {\Large $C_M$};
	\node[draw=none, fill=none, left of= L, yshift=0.2cm, xshift=-1cm](KA) {\Large $C_L$};
	\node[draw=none, fill=none, right of= I, yshift=-0.4cm, xshift=1cm](MA) {\Large $C_I$};
	
	\begin{pgfonlayer}{background}
	\draw[black, fill=gray!15, dashed](E.north) to[closed,curve through={(E.north east) ..  (B.north east) .. (B.east) .. (B.south)  .. (C.south east) .. (C.south west) .. (A.west) .. (E.west)}] (E.north);
	
	\draw[black, fill=gray!15, dashed](M.north) to[closed,curve through={(M.south east) .. (K.south east) .. (K.south west) .. (K.north west)}] (M.north west);
	
	\draw[black, fill=gray!15, dashed](L.north east) to[closed,curve through={(L.south east)(N.south east) .. (N.south west) .. (N.north west)}] (L.north west);
	
	\draw[black, fill=gray!15, dashed] (I.north east) to[closed, curve through={(H.north east) .. (H.east) .. (H.south) .. (G.south) .. (F.south) .. (F.west) .. (F.north west) .. (I.north west)}] (I.north west);
	\end{pgfonlayer}
	
\end{tikzpicture}
}
}}
    \hfill
    \subcaptionbox{Final tree of clusters. \label{fig:stage6}} [0.25\textwidth]
    {\hspace{-1cm} 



\resizebox{0.07\textwidth}{!}
{
\raisebox{2.2cm}{
\begin{tikzpicture}[>=latex',line join=bevel,every fit/.style={regular polygon, regular polygon sides=3,thick, dashed, draw,inner sep=2pt,text width=2cm}]

	\node[draw=gray,fill=black,text=white,circle, yshift=1cm](L_p) {$C_E$};
	
	\node[draw=gray,fill=black,text=white,circle, below =1cm of L_p](C) {$C_M$};
	\node[draw=gray,fill=black,text=white,circle, below =of C](D_p) {$C_L$};
	\node[draw=gray,fill=black,text=white,circle, below =of D_p](F_p) {$C_I$};
	\draw[ultra thick, draw=black] (L_p) -- (C) -- (D_p) -- (F_p);
\end{tikzpicture}
}
}}
    
    \caption{Discovery of clusters with E as the starting node.}
    \label{fig:dfsbfs2}
\end{figure}


From the illustrations, it can be observed that initially a lot of broker nodes are obtained. This is because, at the beginning of the spread propagation inside a community, most of the nodes within that community are unvisited. Therefore, in this method, some small-sized (possibly singleton) communities may be generated at the initial stage. But as the brokers are placed and the community labels of the nodes are stabilized, the communities grow larger and we get the final labels for all the nodes (shown as the final labels in Table~\ref{tab:example1}). Similarly, the initial split can also be generated by using conductance. After applying modularity maximization on the obtained cover, the average size of the clusters grows further.

\subsection{Effect of different starting points on the clusters}

Due to the nature of the algorithm, we can pick any node to be the starting point for the traversal. Depending on the starting point, the order of traversal may change and the broker node that leads to the discovery of a community may change but discovery of the communities remain more or less similar. We have already provided an example in Figure~\ref{fig:cd1}. If other starting points are used on the same network, same cover is produced every time. Figure~\ref{fig:stage5} shows the traversal tree for starting node E, with back edges and cross edges in addition to the tree edges. In Figure~\ref{fig:stage6}, only the tree edges have been shown along with the clusters after the first phase of the algorithm. In this figure, node B is actually connected to cluster $C_E$ with a back edge. After modularity maximization, node B moves into $C_E$, $C_M$ and $C_L$ get merged and $C_I$ remains as it is. This leads to the same final cover as in Figure~\ref{fig:cd1}. In experimental results part, we have provided sufficient empirical evidence that our method is independent of starting point.

\section{Experimental Results}

We have performed our experiments, including running the public releases of the Louvain method and CNM, on an Intel Xeon 2.4 GHz quad-core CPU desktop with 32GB RAM, 500 GB hard disk and Fedora LINUX version 3.3.4 OS. The source code has been written in C and it is publicly available\footnote{The code can be downloaded from \href{https://github.com/sna-lincom/LINCOM}{https://github.com/sna-lincom/LINCOM}}.

We evaluate the performance of our algorithm on different well-known benchmark datasets~\citep{Zachar77,LuNe04,GirvanandNewman2002,YangL12,KYANG04,LeskovecKF07} by assessing the accuracy of its covers obtained, using a Newman modularity for the disjoint case and Nicosia modularity for the overlapping case. We further test our algorithm on large datasets to test its efficiency. We evaluate the effect of parameters used in our algorithm, i.e., threshold values for INS and choice of the starting nodes. The threshold value $r$ for the INS value has a great impact on the size of the communities obtained. If $r$ is assigned a low value, many nodes move to the same community during traversal, leading to mostly large sized communities and only a few smaller ones. On the other hand, taking high value for $r$ leads to fragmented and small communities along with many broker nodes, specially in the initial part of the traversal. A large number of broker nodes may converge to a single community in the latter part of our method, thereby forming large-sized communities. We have observed that setting $r$ = 0.75 as threshold value gives us appropriately sized (with significant relevance to ground truth) communities and therefore in other experiments, INS-based LINCOM was run with $r$ = 0.75.
\begin{table*}[ht]
\centering
\vspace{0.5cm}
\caption [Large benchmark datasets used for experiment and comparing the running time for different types of community detection algorithms]{Comparison of running time for different types of community detection algorithms for different benchmark datasets.}
\ra{1.2}
\begin{tabular}{@{}lrrcrrrr@{}}
\toprule
\multirow{3}{*}{Network Dataset} & \multirow{3}{*}{Nodes} & \multirow{3}{*}{Edges} & \phantom{a} & \multicolumn{2}{c}{LINCOM} & \multirow{3}{*}{Louvain} & \multirow{3}{*}{CNM} \\
\cmidrule{5-6}
 &  & & &  INS &  COND &  &   \\
 & ($n$) & ($m$) & & (sec) &  (sec) & (sec) & (sec)  \\
\midrule
Karate & 34 & 78 & & 0 & 0 & 0 & 0 \\

Dolphin & 62 & 159 & & 0 & 0 & 0 & 0  \\

Lesmis & 77 & 254 & & 0 & 0 & 0 & 0 \\

Football & 115 & 613 & & 0 & 0 & 0 & 0 \\

GrQc & 4,158 & 13,422 & & 0 & 0 & 0 & 4 \\

Enron & 33,696 & 180,811 & & 0.21 & 0.22 & 0.38 & 362 \\

Epinions & 75,877 & 405,739 & & 0.71 & 0.76 & 0.97 & 1953 \\

Amazon & 334,863 & 925,872 & & 4.58 & 4 & 6 & 3578 \\

DBLP & 317,080 & 1,049,866 & & 5.141 & 4 & 11 & 10,440 \\

Orkut & 3,072,441 & 117,185,083 & & 246 & 377 & 456 & - \\
\bottomrule
\end{tabular}
\label{tab:time}
\end{table*}


\begin{table*}[ht]
\centering
\caption [Running time for different types of community detection algorithms]{Comparing running times of the state-of-the-art community detection algorithms with LINCOM for different benchmark datasets. Boldfaced numbers highlight the best time obtained for a given dataset.}
\ra{1.2}
\begin{tabular}{@{}lrrrrrrrrr@{}}
	\toprule  
	 Method  &  Amazon    &    DBLP &   Dolphins &  Enron &  Epinion &  Football &  GrQc &  Karate &  LesMis \\
	\midrule
	Louvain & \textbf{3.46}&    4.36 &       0 &   0.28 &     \textbf{0.73} &       0 &  02 &     0 &     0 \\  
    LexDFS &   13.02     &   14.18 &         0 &   3.40 &    11.24 &       0 &  0.10 &     0 &     0 \\ 
	Label Prop &   27.72    &   43.87 &         0 &   0.69 &     0.93 &       0 &  03 &     0 &     0 \\  
	Infomap &  432.71     &  507.85 &         0 &  22.08 &   110.30 &       0 &  0.54 &     0 &     0 \\  
    LINCOM &     4.00      &   \textbf{4.00} & 0 &    \textbf{0.22} &   0.76  &     0 &    \textbf{0}  &     0  &    0 \\
	\bottomrule
\end{tabular}
\label{tab:time2}
\end{table*}

\subsection{Testing efficiency of LINCOM}

We run INS-based LINCOM and COND-based LINCOM along with the implementation of the Louvain method and CNM, released by their respective authors for a comparative analysis, using the same set-up declared above. The time taken to run those algorithms have been summarized in Table~\ref{tab:time}. Results show that both our methods run faster than the Louvain method. Clearly, CNM proves to be much slower, which is in coherence to the provable bounds for the running time of the algorithm. CNM cannot even find communities for massive datasets such as Orkut even in several hours. So, from the results, it can be empirically established that the concept of traversing the graph to detect communities turned out to be faster than the present state-of-the-art. The variation of the objective function did not seem to generate much perturbation in terms of running times. Hence, other objective functions suitable to maximize intra-cluster edges and minimize inter-cluster edges, can also be tested with similar efficiency unless the objective function itself is computationally expensive.

We have used CODACOM platform~\cite{creusefond2017} to run all the state-of-the-art community detection algorithms and have tested their performance on the real-world benchmark datasets and the synthetic graph data originally used by Newman and Girvan~\cite{GirvanandNewman2002}. In Table~\ref{tab:time2}, we have compared LINCOM with other methods in terms of efficiency using real-world benchmark datasets. In Table~\ref{tab:lfr}, we have shown the results of running LINCOM and a few other state-of-the-art community detection algorithms on synthetic datasets. We created these synthetic datasets using the LFR benchmark generator~\cite{LaFoRa08}. These graphs have 128 nodes with each node having degree of 16. All the nodes in the network is divided into 4 communities with each community consisting of 32 nodes. In these graphs, mixing parameter ($\mu$) defines the fraction of edges with nodes outside its own community. We have changed the value of $\mu$ from 0.1 to 0.3 to test the robustness of our algorithm as the communities become less recognizable with increase in $\mu$.

\begin{table*}[ht]
\centering
\vspace{0.5cm}
\caption [Large benchmark datasets used for experiment and comparing the quality of clusters generated by the community detection algorithms]{Comparison of the modularity values of the final covers generated by the Louvain method, CNM and the variants of LINCOM, for different benchmark datasets. }
\ra{1.2}
\begin{tabular}{@{}lrrcrrrr@{}}
\toprule
\multirow{3}{*}{Network Dataset} & \multicolumn{2}{c}{Overlapping} &\phantom{abc} & \multicolumn{4}{c}{Disjoint}  \\
\cmidrule{2-3} \cmidrule{5-8}
 & LINCOM & LINCOM & & LINCOM & LINCOM & Louvain &  CNM \\
 &  INS &  COND &  & INS &  COND & &  \\
\midrule
Karate & 0.729 & 0.445 & & 0.402 & 0.3793 & 0.415 & 0.38 \\

Dolphin & 0.75 & 0.756 & & 0.518 & 0.526799 & 0.518 & 0.492  \\

LesMis & 0.579 & 0.579 & & 0.544 & 0.448 & 0.55 & 0.5 \\

Football & 0.673 & 0.694 & & 0.582 & 0.543 & 0.604 & 0.57 \\

GrQc & 0.879 & 0.87 & & 0.847 & 0.841375 & 0.847 & 0.79 \\

Enron & 0.73 & 0.7 & & 0.587 & 0.598 & 0.596 & 0.49 \\

Epinions & 0.529 & 0.51 & & 0.44 & 0.44895 & 0.45 & 0.385 \\

Amazon & 0.931 & 0.963 & & 0.961 & 0.92054 & 0.926 & 0.87 \\

DBLP & 0.831 & 0.819 & & 0.818 & 0.809 & 0.819 & 0.73 \\

Orkut & 0.59  & 0.801375 & & 0.548 & 0.679 & 0.679 & - \\
\bottomrule
\end{tabular}
\label{tab:mod}
\end{table*}

\begin{table*}[ht]
\centering
\caption [Modularity values for different types of community detection algorithms]{Modularity values of the final cover of the state-of-the-art community detection algorithms for different benchmark datasets. Boldfaced numbers highlight the highest values of modularity obtained for a given dataset.}
\ra{1.2}
\begin{tabular}{@{}lrrrrrrrrr@{}}
	\toprule  
	Method &  Amazon &    DBLP &  GrQc &  Enron &  Epinion &  Football &  Dolphins &  Karate &  LesMis \\
	\midrule 
	Louvain &  0.926  &    \textbf{0.821} &    0.847 &   \textbf{0.613} & \textbf{0.452} & \textbf{0.6} & 0.518 &     0.392 &  \textbf{0.554} \\ 
	LexDFS &   0.536 &    0.414 &    0.552 &   0.171 &    0.094 &      0.576 &  0.338 &     0.23 &   0.434 \\ 
	Label Prop &   0.785 &    0.7  &    0.769 &   0.317 &     0.044 &     0.584 &  0.410 &     0.352 &  0.523 \\ 
	Infomap &  0.825 &     0.722 &    0.771 &   0.511 &    0.347 &       0.6 &    0.517 &     0.402 &  0.546 \\ 
    LINCOM &     \textbf{0.961}  &    0.818  &    \textbf{0.847} &   0.596 &   0.44  &     0.582 &  \textbf{0.518}  & \textbf{0.402} & 0.544 \\  
	\bottomrule
\end{tabular}
\label{tab:mod2}
\end{table*}

\begin{table*}[ht]
\centering
\vspace{0.5cm}
\caption [Modularity values for different types of community detection algorithms]{Performance of the state-of-the-art community detection algorithms on LFR generated synthetic benchmark datasets.}
\ra{1.2}
\begin{tabular}{@{}lrrr@{}}
	\toprule
	\multirow{2}{*}{Method} & \multicolumn{3}{c}{$\mu$} \\ \cmidrule{2-4}
	 & 0.1 & 0.2 &  0.3 \\
	\midrule 
    Ground truth & 0.648 & 0.548  &  0.452 \\  
	Louvain &  0.648  &    0.548 &   0.452 \\  
	 LexDFS &  0.648 &    0.548 &    0.328 \\  
	Label Prop &  0.648 &    0.548  &    0.230 \\  
	Infomap &  0.648 &     0.548 &    0.452 \\  
    LINCOM &  0.648  &    0.545  &    0.452 \\  
    CNM  & 0.648 &     0.548 &    0.452 \\ \bottomrule
\end{tabular}
\label{tab:lfr}
\end{table*}

\begin{figure*}[ht]
    \centering
    \begin{tikzpicture}
\begin{groupplot}[
    width=0.4\textwidth,
    group style={group size=2 by 1,
        horizontal sep=10pt,
        xlabels at=edge bottom,xticklabels at=edge bottom,ylabels at=edge left,yticklabels at=edge left,},
    legend style={font=\small},
    every tick label/.append style={font=\footnotesize},
    ylabel near ticks,
    xlabel near ticks,
    y label style={at={(axis description cs:0.12,.5)}},
    title style={yshift=-1.0ex,},
	xlabel={{$\%$ Edges}}, 
	ylabel={{$\%$ Running time}},
	xmin=20,
	xmax=105,
	ymin=5,
    ymax=105,
    xtick={25, 50, 75, 100},
    ytick={25, 50, 75, 100},
    ylabel near ticks,
]

\nextgroupplot[title={Without MOD-MAXIMIZE},legend to name={CommonLegend},legend style={legend columns=1}]
\addplot
[ 
opacity=0.8,
black,
dashed,
thick,
mark=oplus*,
mark options={solid, 
draw=black, 
fill=black!10},
]
coordinates {
(25, 25) +- (0, 0)
(50, 50) +- (0, 0)
(75, 75) +- (0, 0)
(100, 100) +- (0, 0)
};

\addplot
[ 
opacity=0.8,
black!80!black,
dashed,
thick,
mark=triangle*,
mark options={solid, 
draw=black, 
fill=black!10},
]
coordinates {
(25, 25.63739377) +- (0, 0)
(50, 49.00849858) +- (0, 0)
(75, 75.07082153) +- (0, 0)
(100, 100) +- (0, 0)
};

\addplot
[ 
opacity=0.7,
black,
dashed,
thick,
mark=square*,
mark options={solid, draw=black, fill=black!10},
]
coordinates {
(25, 25.17810599) +- (0, 0)
(50, 50.28670721) +- (0, 0)
(75, 73.17984361) +- (0, 0)
(100, 100) +- (0, 0)
};

\addplot
[ 
opacity=0.8,
black!40,
dashed,
thick,
mark=diamond*,
mark options={solid, draw=black!40, fill=black!10},
]
coordinates {
(25, 22.14139763) +- (0, 0)
(50, 45.72946465) +- (0, 0)
(75, 68.84348181) +- (0, 0)
(100, 100) +- (0, 0)
};

\addlegendentry{Baseline}
\addlegendentry{Enron}
\addlegendentry{Amazon}
\addlegendentry{DBLP}

\nextgroupplot[title={With MOD-MAXIMIZE},]
\addplot
[ 
opacity=0.8,
black,
dashed,
thick,
mark=oplus*,
mark options={solid, 
draw=black, 
fill=black!10},
]
coordinates {
(25, 25) +- (0, 0)
(50, 50) +- (0, 0)
(75, 75) +- (0, 0)
(100, 100) +- (0, 0)
};
\addplot
[ 
opacity=0.8,
black!80,
thick,
dashed,
mark=triangle*,
mark options={solid, 
draw=black, 
fill=black!10},
]
coordinates {
(25, 24.52941176) +- (0, 0)
(50, 46.64705882) +- (0, 0)
(75, 79.05882353) +- (0, 0)
(100, 100) +- (0, 0)
};

\addplot
[ 
opacity=0.7,
black,
thick,
dashed,
mark=square*,
mark options={solid, draw=black, fill=black!10},
]
coordinates {
(25, 11.46096816) +- (0, 0)
(50, 27.3964239) +- (0, 0)
(75, 50.81988661) +- (0, 0)
(100, 100) +- (0, 0)
};

\addplot
[ 
opacity=0.8,
black!40,
thick,
dashed,
mark=diamond*,
mark options={solid, draw=black!40, fill=black!10},
]
coordinates {
(25, 12.09721678) +- (0, 0)
(50, 31.30929047) +- (0, 0)
(75, 57.69371488) +- (0, 0)
(100, 100) +- (0, 0)
};

\end{groupplot}

\path (group c2r1.north east) -- node[right]{\ref{CommonLegend}} (group c2r1.south east);

\end{tikzpicture}
    \caption[]{
    Growth of running time for LINCOM when tested on 25\%, 50\% and 75\% of the edges of Enron, Amazon and DBLP datasets.
    }
    \label{fig:linearity}
\end{figure*}
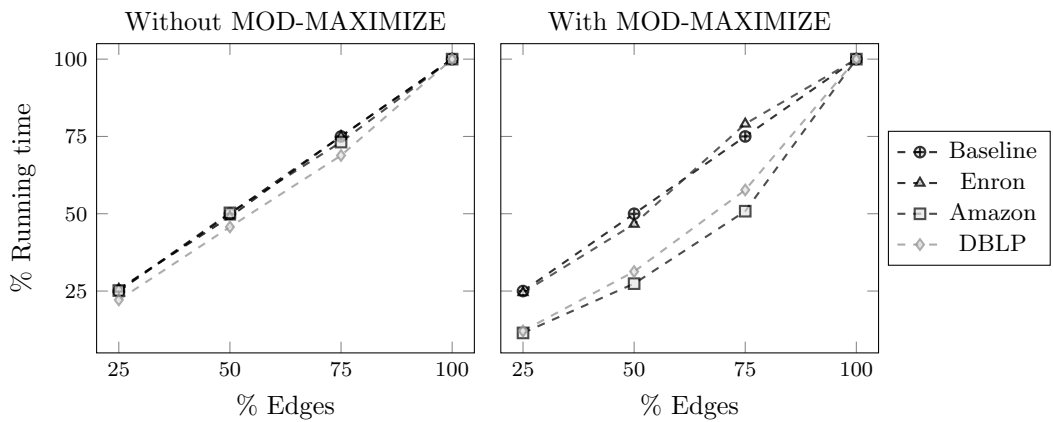

In order to reinforce our claim that the first phase of LINCOM (i.e., LINCOM without modularity maximization) runs in linear time, we have tested LINCOM on sample subgraphs of large datasets by randomly sampling 25\%, 50\% and 75\% of the total edges of the networks. When we use LINCOM with modularity maximization, it shows non-linear growth in running time. If we run LINCOM without the modularity maximization part, results show that the running times scale up linearly. The results have been shown in Figure~\ref{fig:linearity}. This can be considered as an empirical confirmation of the fact that the running time of LINCOM is indeed bounded by a linear function of the number of edges in the network.

\subsection{Testing Quality of the Clusters}

Here, we have used overlapping modularity~\citep{NPNB07,HWXJ09,Nicosia2009} for evaluating overlapping communities, whereas for disjoint communities, we use modularity defined by Newman~\citep{Newman06}. The variants of LINCOM are naturally overlapping, however, in order to compare them with disjoint communities on the basis of one goodness measure, we convert the overlapping communities into disjoint communities. We place each overlapping node in one of its neighboring communities such that the modularity is maximized. A summary of the results have been presented in Table~\ref{tab:mod}. Overlapping modularity of the covers generated by the variants of LINCOM are consistently greater than the modularity values of the covers produced by the Louvain method and CNM. The comparisons between the modularity values of the disjoint covers generated by the variants of LINCOM and the modularity maximization methods are close. Variants of LINCOM always seem to produce better clusters than that of CNM, particularly in large networks. The Louvain method and variants of LINCOM often produce covers with the same modularity value with a very low tolerance level.

In Table~\ref{tab:mod2}, we have compared performance of LINCOM with the performance of some of the state-of-the-art community detection algorithms. LINCOM consistently performs better than the others except the Louvain method. Louvain method outputs covers with similar modularity values in some cases. In Table~\ref{tab:lfr}, we have evaluated LINCOM's performance on synthetic networks with 4 existing communities where the community structure becomes more obscure as the mixing factor increases. The ground truth is available and the modularity of the ground truth cover decreases as the mixing factor increases. From the results, LINCOM seems to be more robust than some of the other methods as it can recognize the ground truth structure more readily than the others.

\begin{figure}[htb]
    \centering
    \vspace{0.5cm}
    \includegraphics[scale=0.24]{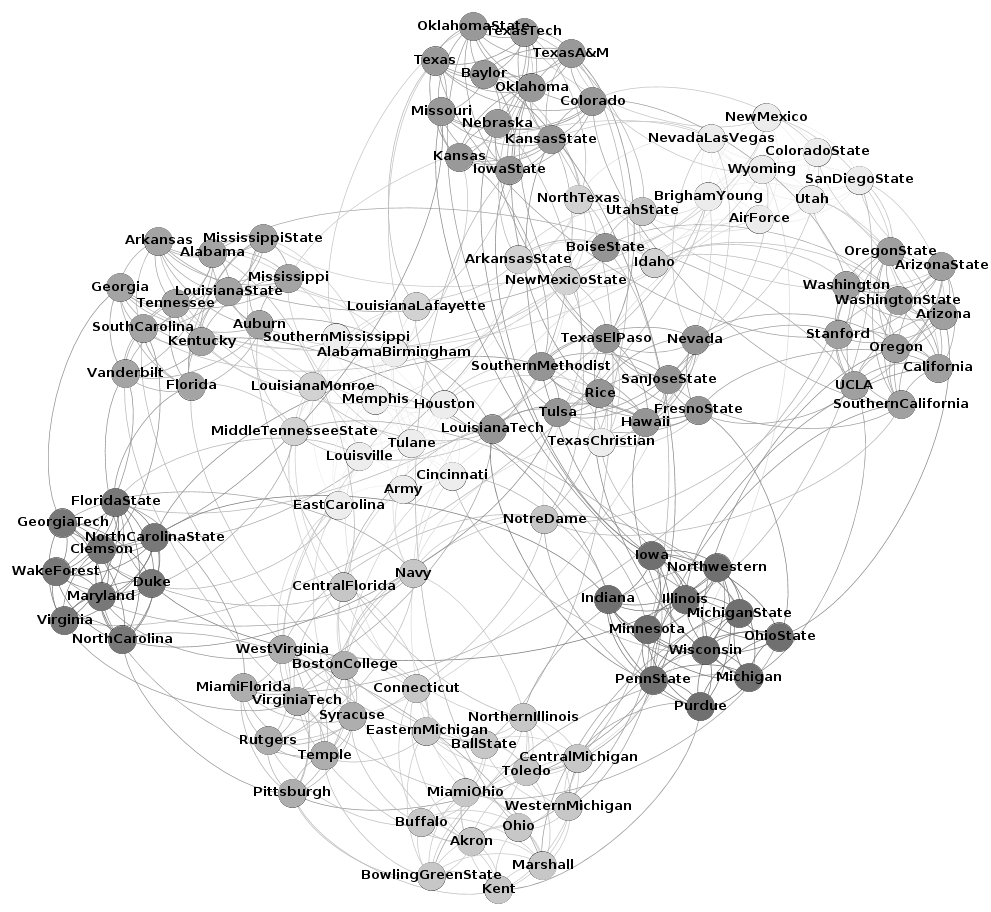}
    \caption[]{
    Ground-truth for the colleges in the twelve conferences~\citep{GirvanandNewman2002}.
    }
    \label{fig:ground}
\end{figure}

\begin{figure*}[ht]
    \vspace{0.5cm}
    \subcaptionbox{Final cover obtained from Louvain method. \label{fig:coverLouvain}}[0.50\textwidth]
    {\includegraphics[width=0.5\textwidth]{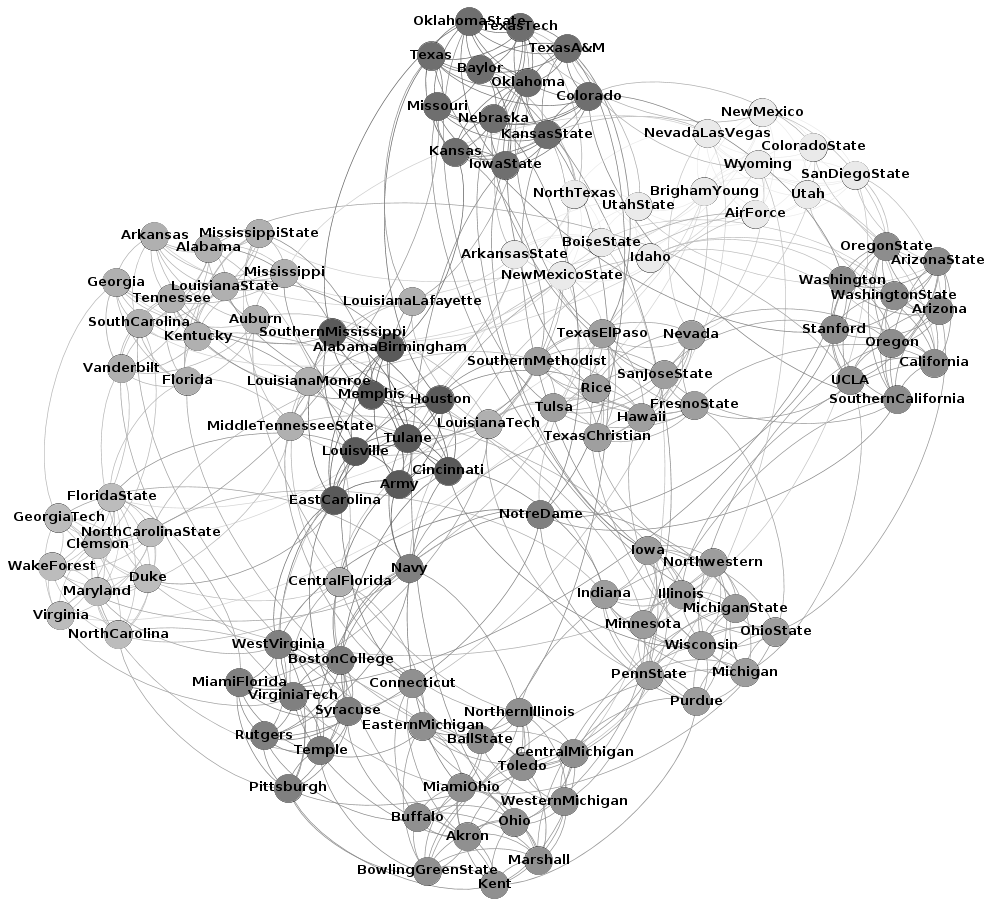}}
    \hfill
    \subcaptionbox{Final cover obtained from LINCOM. \label{fig:coverLINCOM}} [0.5\textwidth]
    {\includegraphics[width=0.5\textwidth]{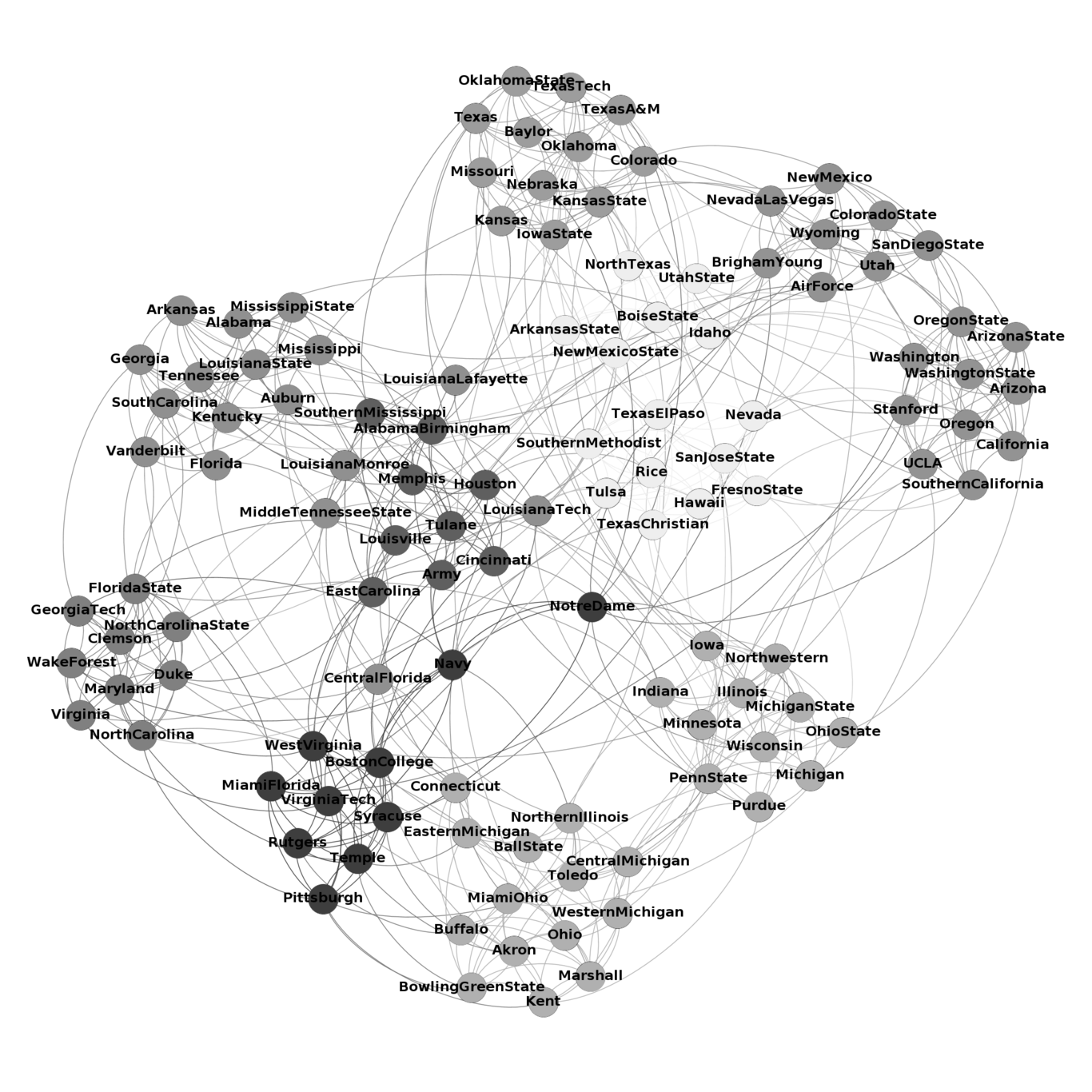}}
    
    \caption{Visualization of the final covers as obtained from Louvain method and LINCOM.}

    \label{fig:sig}
\end{figure*}

In every community detection algorithm, it is important to understand the practical significance of the method and what it can be used for. To understand that we compare the final cover obtained by INS-based LINCOM method (Figure~\ref{fig:coverLINCOM}) with the ground-truth of the US College football network as described by Newman~\citep{GirvanandNewman2002} (Figure~\ref{fig:ground}) and the final cover obtained by Louvain method (Figure~\ref{fig:coverLouvain}). This is one of the very few datasets where ground-truth from real-world is available and therefore can be used for benchmarking communities. We observe that Louvain generates a cover with ten clusters whereas LINCOM ends up generating a cover with eight clusters. Interestingly, the significance of the ground-truth communities is maintained both in Louvain as well as LINCOM. In the final cover generated by Louvain method we can observe that there are two cases where a couple of conferences have merged. Similarly, a further degree of coarsening has taken place in LINCOM, where two pairs of communities have merged to reduce the number of clusters further. Understandably, due to LINCOM's use of modularity maximization technique, inability to detect small clusters has become an inherent problem. But the clustering found in LINCOM is defined by dense connection, a higher modularity value and evidently has a high precision when compared to the ground truth.

\subsection{Selection of Starting Node in LINCOM}

\begin{figure*}[htb]
    \centering
    \vspace{0.75cm}
    \begin{tikzpicture}
    \begin{groupplot}[
    width=2.3in,
    group style={group size=3 by 2,
        horizontal sep=15pt,
        vertical sep=70pt,
        ylabels at=edge left,
        yticklabels at=edge left,},
    log ticks with fixed point,
	xlabel={Modularity $(Q)$},
	ylabel={Centrality},
	ymode=log,
    ymax=1,
    ymin=0.00001,
    xmin=0,
    xmax=1,
    x tick label style={
    /pgf/number format/.cd,
    precision=3
  }
]
\pgfplotstableread{./data/karate}
\karate

\pgfplotstableread{./data/dolphins}
\dolphins

\pgfplotstableread{./data/football}
\football

\pgfplotstableread{./data/grqc}
\grqc

\pgfplotstableread{./data/facebook}
\facebook

\pgfplotstableread{./data/enron}
\enron

\nextgroupplot[title={Karate \\ $(n = 34,\ m = 78)$}, align=center]

\addplot
[
opacity=0.6,
mark=diamond,
only marks,
draw=black,
]
table[y = DC] from \karate;

\addplot
[
opacity=0.6,
mark=square,
only marks,
draw=black,
]
table[y = PR] from \karate;

\addplot
[
opacity=0.6,
mark=o,
only marks,
draw=black,
]
table[y = BC] from \karate;

\addplot
[
opacity=0.6,
mark=triangle,
only marks,
draw=black,
]
table[y = CC] from \karate;

\nextgroupplot[title={Dolphin \\ $(n = 62,\ m =
159)$},
align=center]

\addplot
[
opacity=0.6,
mark=diamond,
only marks,
draw=black,
]
table[y = DC] from \dolphins;

\addplot
[
opacity=0.6,
mark=square,
only marks,
draw=black,
]
table[y = PR] from \dolphins;

\addplot
[
opacity=0.6,
mark=o,
only marks,
draw=black,
]
table[y = BC] from \dolphins;

\addplot
[
opacity=0.6,
mark=triangle,
only marks,
draw=black,
]
table[y = CC] from \dolphins;



\nextgroupplot[title={US Football \\ $(n = 115,\
 m = 613)$}, align=center]

\addplot
[
opacity=0.6,
mark=diamond,
only marks,
draw=black,
]
table[y = DC] from \football;

\addplot
[
opacity=0.6,
mark=square,
only marks,
draw=black,
]
table[y = PR] from \football;

\addplot
[
opacity=0.6,
mark=o,
only marks,
draw=black,
]
table[y = BC] from \football;

\addplot
[
opacity=0.6,
mark=triangle,
only marks,
draw=black,
]
table[y = CC] from \football;

\nextgroupplot[title={GrQc \\ $ (n = 4,158,\ m
= 13,422)$}, 
align=center]

\addplot
[
opacity=0.6,
mark=diamond,
only marks,
draw=black,
]
table[y = DC] from \grqc;

\addplot
[
opacity=0.6,
mark=square,
only marks,
draw=black,
]
table[y = PR] from \grqc;

\addplot
[
opacity=0.6,
mark=o,
only marks,
draw=black,
]
table[y = BC] from \grqc;

\addplot
[
opacity=0.6,
mark=triangle,
only marks,
draw=black,
]
table[y = CC] from \grqc;

\nextgroupplot[title={Facebook \\ $(n = 4,039,\ m
= 88,234)$}, align=center]

\addplot
[
opacity=0.6,
mark=diamond,
only marks,
draw=black,
]
table[y = DC] from \facebook;

\addplot
[
opacity=0.6,
mark=square,
only marks,
draw=black,
]
table[y = PR] from \facebook;

\addplot
[
opacity=0.6,
mark=o,
only marks,
draw=black,
]
table[y = BC] from \facebook;

\addplot
[
opacity=0.6,
mark=triangle,
only marks,
draw=black,
]
table[y = CC] from \facebook;

\nextgroupplot[title={Enron \\ $(n = 33,696,\ m
= 180,811)$}, align=center]

\addplot
[
opacity=0.6,
mark=diamond,
only marks,
draw=black,
]
table[y = DC] from \enron;

\addplot
[
opacity=0.6,
mark=square,
only marks,
draw=black,
]
table[y = PR] from \enron;

\addplot
[
opacity=0.6,
mark=o,
only marks,
draw=black,
]
table[y = BC] from \enron;

\addplot
[
opacity=0.6,
mark=triangle,
only marks,
draw=black,
]
table[y = CC] from \enron;

\end{groupplot}
\end{tikzpicture}
    \begin{tikzpicture}
    \begin{customlegend}[ 
    legend columns=4,
    legend style={
    draw=none,
    column sep=1ex,
	anchor=north
  },
  legend entries={Degree, PageRank, Betweenness, Closeness}]

    \addlegendimage{black, only marks, mark=diamond}
    \addlegendimage{black, only marks, mark=square}
    \addlegendimage{black, only marks, mark=o}
    \addlegendimage{black, only marks, mark=triangle}
    
    \end{customlegend}
\end{tikzpicture}
    \caption{
    For the networks in the first row, every node in the network was sequentially tested as the starting node. For the networks in the bottom row, $500$ nodes were randomly sampled and then their centrality data is plotted. The modularity of the final cover found using that starting node was plotted against its degree centrality, PageRank, betweenness centrality and closeness centrality.
    }
    \label{fig:startNode}
\end{figure*}
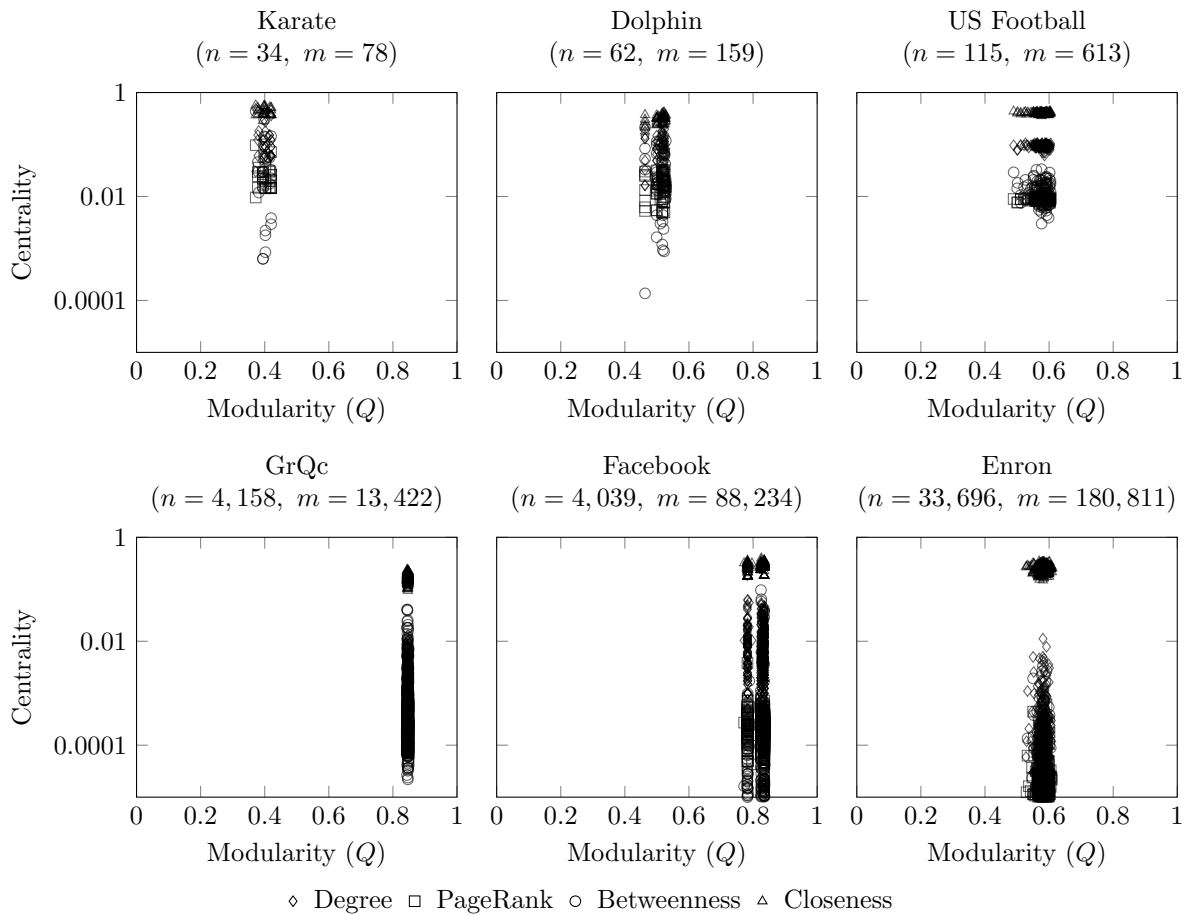

Starting points may differ in traversal methods and we observed that LINCOM generated different covers from different starting nodes. We used INS-based LINCOM to generate covers exhaustively by using every node of the network as starting points. For large networks this process proved to be time-consuming. In order to solve that problem, for large networks, we chose a random sample of starting nodes to figure out a similar statistics. We plotted the centrality measures of the starting node, such as degree centrality, PageRank, betweenness centrality and closeness centrality against the modularity of the final cover obtained. We observed that the modularity does not seem correlated to the centrality measures mentioned. From the plots in Figure~\ref{fig:startNode}, we can clearly state that a good starting node cannot be selected by the usual network features of a node such as centrality measures. On the other hand, the experiment uncovered an interesting fact that whatever the starting node might be, the modularity does not change significantly. From experiments, we found out (see Table~\ref{tab:startMod}) that we can randomly pick any node in the network as a starting node to find a cover and still guarantee that the modularity of the cover will be within a very small relative standard deviation (usually 0-3\%) of the mean modularity value of all the covers generated by using every node of the network as starting point.

\begin{table*}[ht]
\centering
\vspace{0.5cm}

\caption []{Comparison of the reported modularity values of the final covers generated by LINCOM, and the mean modularity value of all the covers generated by using every node in the network as starting point.}
\ra{1.2}
\begin{tabular}{@{}lrrr@{}}
\toprule
\multirow{2}{*}{Network Dataset} & Mean & Standard & Reported  \\
 & Modularity & Deviation & Modularity  \\
\midrule
Karate & 0.402 & 0.015 & 0.402 \\

Dolphin & 0.518 & 0.014 & 0.518 \\

Lesmis & 0.544 & 0.011 & 0.544 \\

Football & 0.581 & 0.017 & 0.582 \\

GrQc & 0.847 & 0.001 & 0.847 \\

Enron & 0.578 & 0.013 & 0.587 \\

Facebook & 0.818  & 0.023 & 0.835 \\
\bottomrule
\end{tabular}
\label{tab:startMod}
\end{table*}

\subsection{Selection of INS threshold (\texorpdfstring{$r$}{r})}

\begin{table*}[ht]
\centering
\vspace{0.5cm}
\caption []{Comparison of the modularity values of the final covers generated by the INS-based LINCOM, by varying the value of INS threshold ($r$), for different benchmark datasets. Boldfaced numbers highlight the highest values of modularity obtained for a given dataset.}
\ra{1.2}
\begin{tabular}{@{}lrrrrrrrrr@{}}
\toprule
Threshold (r) & Karate & Dolphin & Football & LesMis & Facebook & GrQc & Enron & Amazon & DBLP \\
\midrule
0.4 & 0     & 0.475         & 0             & 0.331         & 0.651         & 0.765             & 0.574             & 0.959  & 0.70  \\

0.45 & 0    & 0.483         & 0             & 0.408         & 0.700         & 0.822             & 0.574         & \textbf{0.961}  & 0.74 \\

0.5 & 0     & 0.480         & 0             & 0.282         & 0.776         & 0.832             & 0.582         & \textbf{0.961}  & 0.78  \\

0.55 & 0.372 & \textbf{0.526} & 0.428       & 0.504          & 0.826        & 0.840             & 0.598          & \textbf{0.961}  & 0.80  \\

0.6 & 0.372  & 0.518        & 0.435         & 0.529         & 0.821         & 0.843             & \textbf{0.604}  & \textbf{0.961}  & 0.81  \\

0.65 & 0.372 & \textbf{0.526} & 0.541       & 0.529         & 0.832         & 0.842             & 0.581          & \textbf{0.961}  & \textbf{0.82}  \\

0.7 & 0.372  & \textbf{0.526} & 0.541       & 0.529         & \textbf{0.835} & 0.843            & 0.602         & \textbf{0.961}  & \textbf{0.82}  \\

0.75 & 0.372 & \textbf{0.526} & \textbf{0.6}   & 0.521      & \textbf{0.835} & \textbf{0.847} & 0.575           & \textbf{0.961}  & \textbf{0.82}  \\

0.8 & \textbf{0.42}   & 0.525 & 0.586       & 0.529         & \textbf{0.835} & 0.846            & 0.6           & 0.960  & \textbf{0.82}  \\

0.85 & 0.39  & 0.524           & 0.574      & \textbf{0.560} & 0.834         & 0.846            & \textbf{0.604}  & 0.959  & \textbf{0.82}  \\
\bottomrule
\end{tabular}
\label{tab:threshold}
\end{table*}

We have performed experiments to run INS-based LINCOM over several INS threshold values, thereby creating a variety of final covers (as shown in Table~\ref{tab:threshold}). We compared the quality of those final covers for each dataset and found that most of the benchmark datasets produce the best cover when INS threshold ($r$) is 0.75. Therefore, in other experiments, we have chosen to keep the value of $r$ as 0.75.

\subsection{Size of Communities produced by LINCOM}

\begin{table}[ht!]
\centering
\vspace{0.5cm}
\caption [Large benchmark datasets used for experiment and comparing the cover size for different types of community detection algorithms]{Comparison of cover size for different types of community detection algorithms for different benchmark datasets.}
\ra{1.2}
\begin{tabular}{@{}lrrrr@{}}
\toprule
\multirow{3}{*}{Network Dataset} & \multicolumn{2}{c}{LINCOM} & \multirow{3}{*}{Louvain} & \multirow{3}{*}{CNM}  \\ \cmidrule{2-3}
 & INS & COND &  &   \\
  & $|G_s|$ & $|G_s|$ & $|G_s|$ & $|G_s|$ \\
\midrule
Karate & 2 & 3 & 4 & 3  \\

Dolphin & 3 & 2 & 5 & 4   \\

Lesmis & 2 & 2 & 6 & 5  \\

Football & 6 & 5 & 9 & 7  \\

GrQc &  28 & 13 & 42 & 61 \\

Enron & 61 & 23 & 170 & 567 \\

Epinions & 671 & 541 & 733 & 2,983 \\

Amazon & 121 & 59 & 246 & 1,409 \\

DBLP & 101 & 228 & 244 & 3,113 \\

Orkut & 136 & 6 & 11 & - \\
\bottomrule
\end{tabular}
\label{tab:gs}
\end{table}

We observe the number of final clusters in each of these methods (as shown in Table~\ref{tab:gs}). In some cases, LINCOM produces covers of similar size, when compared to the Louvain method. In most other cases, the size of the covers generated by our methods are smaller, i.e., the communities are larger compared to the Louvain method. Therefore, LINCOM guarantees that the communities produced are not too small to be considered insignificant. Smaller cover size ensures average size of clusters are larger. A cover with larger sized clusters produce higher modularity only if the community structures are more meaningful than a cover with smaller clusters.
\vspace{-0.25cm}
\section{Conclusion}
We have convincingly shown that efficient detection of \emph{high quality} communities can be achieved in near linear running time (in terms of the size of the graph) by combining simple traversal methods such as breadth first and depth first traversals. Our methods run faster than the state-of-the-art community detection techniques. Also, the quality of the communities generated by the proposed methods are at least as good as the state-of-the-art community detection techniques.
\vspace{-0.25cm}
\section*{Acknowledgement}

We would like to thank the reviewers for their effort for thoroughly reading our paper and for suggesting valuable changes. We would also like to thank Upasana Dutta for pointing out and correcting errors in some of the figures and algorithms that appear in this work. 

\bibliographystyle{plainnat}

\bibliography{ecai2014_pks}





\end{document}